\documentclass[aip, jcp, amsmath,amssymb, reprint, footinbib,floatfix,citeautoscript]{revtex4-1}

\usepackage{graphicx}% Include figure files
\usepackage{dcolumn}% Align table columns on decimal point
\usepackage{bm}% bold math
\usepackage{color}

\usepackage{mathtools}
\usepackage{centernot}
\usepackage{braket}

\begin{document}

\title{On the exact continuous mapping of fermions}

\author{Andr\'es Montoya-Castillo}
\affiliation{Department of Chemistry, Stanford University, Stanford, California, 94305, USA}

\author{Thomas E. Markland}
\email{tmarkland@stanford.edu}
\affiliation{Department of Chemistry, Stanford University, Stanford, California, 94305, USA}

\date{\today}

\begin{abstract}
We derive a rigorous, quantum mechanical map of fermionic creation and annihilation operators to continuous Cartesian variables that exactly reproduces the matrix structure of the many-fermion problem. We show how our scheme can be used to map a general many-fermion Hamiltonian and then consider two specific models that encode the fundamental physics of many fermionic systems, the Anderson impurity and Hubbard models. We use these models to demonstrate how efficient mappings of these Hamiltonians can be constructed using a judicious choice of index ordering of the fermions. This development provides an alternative exact route to calculate the static and dynamical properties of fermionic systems and sets the stage to exploit the quantum-classical and semiclassical hierarchies to systematically derive methods offering a range of accuracies, thus enabling the study of problems where the fermionic degrees of freedom are coupled to complex anharmonic nuclear motion and spins which lie beyond the reach of most currently available methods. 
\end{abstract}

{\maketitle}

\normalsize

The Meyer-Miller-Stock-Thoss (MMST) approach provides an exact prescription to map a Hamiltonian consisting of discrete states to one in terms of continuous Cartesian phase space variables (positions and momenta). Originally introduced as the ``classical electron'' model \cite{Meyer1979}, this approach was later generalized and shown to be a rigorous quantum mechanical representation \cite{Stock1997}. The ability to represent the Hamiltonian in terms of continuous Cartesian phase space variables facilitates the use of classical-like trajectories to obtain quantum mechanical information via exact path integral approaches as well as quantum-classical and semiclassical approximations. In particular, the MMST mapping has provided the cornerstone for the development and application of a large family of nonadiabatic methods based on quantum-classical \cite{Volobuev2000, Bonella2005, Dunkel2008, Huo2011, Kim2008, Hsieh2012, Kelly2012} and semiclassical \cite{Sun1997, Muller1998, Muller1999, Wang1998b, Sun1998, Thoss1999, Thoss2000, Coronado2001, Liao2002, Shi2004a, Ananth2007, Ananth2010, Ananth2013, Richardson2013, Hele2016, Chowdhury2017, Church2018} approximations to the mapped propagator. However, while the MMST protocol provides a route to map Hamiltonians containing discrete states, for those that include fermionic creation and annihilation operators an exact Cartesian mapping has remained elusive. The lack of a mapping approach for fermionic operators has thus prevented the application of similar approaches to many-fermion problems where the discrete energy levels are so numerous as to create continua, as is the case for processes near metallic and semiconducting interfaces.

In principle, one could apply the MMST approach to problems containing fermionic creation and annihilation operators by expressing the second-quantized operators in terms of outer products of the many-body basis. However, the Hilbert space constructed using $M$ single-particle orbitals contains $2^M$ many-body states, which, upon mapping, results in twice this number of phase space variables ($2^{M+1}$). This exponential scaling with the number of single-particle orbitals is mildly ameliorated in cases involving a fixed number of fermions, $N$, where the $2^M$ dimensional Hilbert space can be limited to the $M!/N!(M-N)!$ dimensional Fock space, but in practice this still renders the MMST treatment infeasible in most cases.

To obviate this highly unfavorable scaling with the number of single-particle orbitals, one could instead consider directly mapping the fermionic creation and annihilation operators, which naturally encode the antisymmetry of the fermionic wavefunction by virtue of their anticommutivity. Since this would allow linear scaling in the number of orbitals, a number of continuous mappings have been suggested based either on fermion coherent states \cite{Plimak2001, Corney2004, Corney2006, Corney2006a, Dalton2016, Polyakov2016} or heuristic connections in the classical limit \cite{Miller1978, Remacle2000a, Li2012, Li2014b}. 

Here, we derive a rigorous mapping that provides a general approach to represent fermionic creation and annihilation operators as continuous Cartesian phase space variables. Our map thus provides an exact starting point for the application of the entire arsenal of quantum-classical and semiclassical techniques to investigate the statics and dynamics of problems involving many fermions. These methods can be used to elucidate physical processes in systems ranging from electrochemical interfaces to nanojunctions and strongly correlated materials. 

Our objective is to map a general many-fermion Hamiltonian of the form
\begin{equation} \label{eq:general-fermion-hamiltonian}
    \hat{H} = \sum_{j,k} h_{j,k}(\boldsymbol{\Gamma}) \hat{c}_{j}^{\dagger}\hat{c}_{k} + \frac{1}{2} \sum_{j,k,l,m} U_{jk,lm}(\boldsymbol{\Gamma}) \hat{c}_{j}^{\dagger}\hat{c}_{k}^{\dagger}\hat{c}_{m}\hat{c}_{l},
\end{equation}
where $\hat{c}_{j}^{\dagger}$ and $\hat{c}_{j}$ are the fermionic creation and annihilation operators for the $j^{\mathrm{th}}$ single-particle orbital, $h_{j,k}(\boldsymbol{\Gamma})$ are the matrix elements of the single-particle part of the Hamiltonian and $U_{jk,lm}(\boldsymbol{\Gamma})$ are those of the two-body component. In the general case, these matrix elements could themselves depend on an additional set of continuous or discrete degrees of freedom, $\boldsymbol{\Gamma} \equiv \{\mathbf{P}, \mathbf{Q}, \ket{r}\bra{s} \}$. This type of Hamiltonian encompasses a broad class of systems, including those where the fermionic dynamics are coupled to complex atomistic or spin degrees of freedom. For example, Eq.~(\ref{eq:general-fermion-hamiltonian}) encompasses the Hubbard \cite{Gutzwiller1963, Hubbard1963, Kanamori1963}, Anderson impurity \cite{Anderson1961}, and Holstein \cite{Holstein1959a} models, which form the basis of our description of processes such as superconductivity in correlated materials \cite{Dagotto1994, Alexandrov1994, Lee2006, Kohno2012}, charge conduction in nanoscopic junctions \cite{Cornaglia2004, Werner2007, Han2010}, and polaron formation \cite{Holstein1959b, Devreese2009}.

To introduce an exact continuous mapping of Eq.~(\ref{eq:general-fermion-hamiltonian}), we first express the fermionic operators in terms of two level system (spin 1/2) operators. While spins can describe the occupied and unoccupied states of a single-particle orbital, spins on different sites do not naturally anticommute with each other. Hence, it is necessary to encode the fermionic anticommutivity, which can be done formally by employing the Jordan-Wigner (JW) transformation \cite{Jordan1928, Giamarchi2004}. This transformation has previously been used to enable the solution of fundamental problems in magnetism \cite{Giamarchi2004} and, more recently, the quantum simulation of fermionic Hamiltonians \cite{Ortiz2000, Somma2001, Georgescu2014, Barends2015, Babbush2016}. By using the JW transformation, one can exactly map a set of $M$ second quantized fermionic operators corresponding to the creation or annihilation of a fermion in a single-particle orbital to $M$ spins arranged in a one-dimensional (1D) lattice,
    \begin{subequations} \label{eq:JW-transformation}
        \begin{eqnarray}
    	\hat{c}_j & \mapsto & \hat{F}(0,j) \sigma_j^{-}, \label{eq:JW-transformation-annihilation} \\
    	\hat{c}_j^{\dagger} & \mapsto & \hat{F}(0,j) \sigma_j^{+}, \label{eq:JW-transformation-creation}
    	\end{eqnarray}
    \end{subequations}
where
    \begin{equation}
        \hat{F}(j,k) \equiv \prod_{l=\min[j+1,k+1]}^{\max[j-1,k-1]} f(\sigma_l^{z}) \label{eq:definition-nonlocal-F}
    \end{equation}
is the nonlocal operator that imposes the fermionic anticommutivity, $j,k \in \{ 1, ..., M\}$, and the operators for the $j^{\mathrm{th}}$ spin are 
    \begin{subequations} \label{eq:sigmas}
        \begin{align}
        \sigma_j^+   &= \ket{1_j}\bra{0_j}, \label{eq:sigma+} \\
        \sigma_j^-   &= \ket{0_j}\bra{1_j}, \label{eq:sigma-} \\
        \sigma_j^{z} &= \ket{1_j}\bra{1_j} - \ket{0_j}\bra{0_j}. \label{eq:sigmaz}
        \end{align}
    \end{subequations}
It is often convenient to set $f(\sigma_l^{z}) = -\sigma_l^{z}$ \cite{Ortiz2000}, which trivially yields the correct behavior when the problem is treated quantum mechanically. However, the function $f(\sigma_l^{z})$ in Eq.~(\ref{eq:definition-nonlocal-F}) can be shown to give the exact quantum mechanical solution as long as it outputs $-1$ when the $l^{\mathrm{th}}$ single-particle orbital is occupied and $+1$ otherwise. This is important to note since a different functional form may prove more advantageous if one were to treat the mapped many-fermion Hamiltonian using quantum-classical or semiclassical theories, where $\sigma_l^{z}$ could take values different from $\pm 1$. However, as long as $f(\sigma_l^{z})$ satisfies the above requirement, it is simple to confirm that the JW transformation exactly reproduces the fermionic anticommutation relations, $\{ \hat{c}_j, \hat{c}_k^{\dagger} \} = \{ \hat{F}(0,j) \sigma_j^{-}, \hat{F}(0,k) \sigma_k^{+} \} = \delta_{j,k}$ \cite{Giamarchi2004}.

It is insightful to briefly consider how the JW transformation encodes the fermionic anticommutivity via the ordering of spins along a 1D chain. By arranging the fermionic single-particle orbitals used to construct the many-body Hilbert space along a 1D chain, the JW transformation keeps a record of the normal ordering of the many-body basis. By additionally including the operator, $\hat{F}(j,k)$, which exploits the normal ordering of the many body-basis mirrored in the 1D chain arrangement, the JW transformation imposes anticommutivity on the spins, which otherwise commute. This allows the mapped operators to act on the many-body basis in the same way as the original fermionic creation and annihilation operators.

We now express the discrete spin states resulting from the JW transformation in terms of bosonic degrees of freedom. To achieve this, we employ the Schwinger theory of angular momentum \cite{Schwinger1965} to express spin operators as coupled bosons. This map was initially developed to easily obtain rotation matrices and Clebsch-Gordan coefficients \cite{Schwinger1965} and subsequently exploited in the semiclassical theories of magnetism \cite{Auerbach1998, Mattis1988}. The Schwinger transformation maps a single spin labelled by index $j$ to two coupled bosons, to which we refer as the $\alpha$ and $\beta$ modes,
\begin{subequations} \label{eq:sigmas-schwinger}
    \begin{align}
    \sigma_j^+ \quad &``\mapsto" \quad \hat{b}_{j\beta}^{\dagger}\hat{b}_{j\alpha}, \label{eq:sigma+schwinger} \\
    \sigma_j^- \quad  &``\mapsto" \quad \hat{b}_{j\alpha}^{\dagger}\hat{b}_{j\beta}, \label{eq:sigma-schwinger} \\
    \sigma_j^{z} \quad &``\mapsto" \quad \hat{b}_{j\beta}^{\dagger}\hat{b}_{j\beta} - \hat{b}_{j\alpha}^{\dagger}\hat{b}_{j\alpha}. \label{eq:sigmazschwinger}
    \end{align}
\end{subequations}
Here the quotation marks indicate, as shown in Supplemental Material, Sec.~\ref{sec:schwinger-theory-basics}, that the Schwinger transformation is not exact on an operator level. However, as shown in the Supplemental Material, Sec.~\ref{sec:schwinger-theory-basics}, the map becomes exact when one restricts it to the physical basis, i.e., the joint single-excitation subspace consisting of the states for which the $\alpha$ mode is in its first excited state and the $\beta$ mode is in its ground state, and vice versa, for every $j$. Hence, while the Schwinger theory of angular momentum does not constitute an isomorphism on the operator level, it is an {\it exact} isomorphism on the matrix element level when they are evaluated using the basis consisting of only the single-excitation manifold for each $j$. Indeed, this can be confirmed by using the mapped spin ladder operators, $\sigma_j^+$ and $\sigma_j^-$, to construct the spin polarization and unit operators, $\sigma_j^z = \sigma_j^-\sigma_j^+ - \sigma_j^+\sigma_j^-$ and $\mathbf{1}_j = \sigma_j^-\sigma_j^+ + \sigma_j^+\sigma_j^-$ and noting that the correct form of the latter can be recovered when excitations outside the single-excitation manifold are eliminated. We are now in a position to combine the Schwinger map with the JW transformation to yield an exact bosonic representation of fermionic matrix elements.

As we demonstrate in the Supplemental Material, Sec.~\ref{sec:fermion-to-boson-map}, using Eqs.~(\ref{eq:JW-transformation}), (\ref{eq:definition-nonlocal-F}) and (\ref{eq:sigmas-schwinger}), one can derive an exact isomorphic representation of fermionic matrix elements in terms of bosonic ones,
\begin{equation} \label{eq:arbitrary-operator-map-schwinger-matrix-elements}
    \bra{ \mathbf{\tilde{n}} } \hat{O}(\{\hat{c}_j, \hat{c}_j^{\dagger}\})\ket{\mathbf{\tilde{n}}^{\prime}} \mapsto  \bra{ \mathbf{n} }\hat{O}(\{ \hat{b}_{j\gamma}^{\dagger}, \hat{b}_{j\gamma} \}) \ket{\mathbf{n}^{\prime}}.
\end{equation}
Here, the ordered fermionic occupation number basis, $\tilde{\mathbf{n}} \equiv \{ \tilde{n}_1, \tilde{n}_2, ..., \tilde{n}_M \}$ where $\tilde{n}_j \in \{ 0,1 \}$, is mapped to its bosonic counterpart, $\tilde{\mathbf{n}} \equiv \{ n_{1\alpha}, n_{1\beta}, n_{2\alpha}, n_{2\beta}, ..., n_{M\alpha}, n_{M\beta} \} $, where $n_{j\beta} = \tilde{n}_{j}$ and $n_{j\beta} = 1 - \tilde{n}_{j}$. An arbitrary fermionic operator $\hat{O}(\{ \hat{c}_j, \hat{c}_j^{\dagger}\})$ can then be written in terms of bosonic operators $\hat{O}(\{ \hat{b}_{j\gamma}^{\dagger}, \hat{b}_{j\gamma}\} )$, where $\gamma \in \{ \alpha, \beta \}$, using,
\begin{subequations} \label{eq:JW-transformation-schwinger}
    \begin{align}
    \hat{c}_j & \ ``\mapsto" \ \hat{F}(0,j) \hat{b}_{j\alpha}^{\dagger}\hat{b}_{j\beta}, \label{eq:JW-transformation-annihilation-schwinger} \\
	\hat{c}_j^{\dagger} & \ ``\mapsto" \ \hat{F}(0,j) \hat{b}_{j\beta}^{\dagger}\hat{b}_{j\alpha}, \label{eq:JW-transformation-creation-schwinger}
	\end{align}
\end{subequations}
where
\begin{equation}
    \hat{F}(j,k) \ ``\mapsto" \ \prod_{l=\min[j+1,k+1]}^{\max[j-1,k-1]} f(\hat{b}_{j\beta}^{\dagger}\hat{b}_{j\beta}- \hat{b}_{j\alpha}^{\dagger}\hat{b}_{j\alpha}). \label{eq:definition-nonlocal-F-schwinger}
\end{equation}
Here the quotation marks around the map symbol emphasize that the transformation in Eqs.~(\ref{eq:JW-transformation-schwinger})-(\ref{eq:definition-nonlocal-F-schwinger}) only works on the operator level if one eliminates all the excitations that lie outside of the physical subspace of the $\alpha$ and $\beta$ modes, which one can exactly enforce by using the physical basis (Supplemental Material, Sec.~\ref{sec:fermion-to-boson-map}). Equations (\ref{eq:arbitrary-operator-map-schwinger-matrix-elements})-(\ref{eq:definition-nonlocal-F-schwinger}) thus provide a formal prescription to exactly obtain the matrix elements of fermionic operators from an isomorphic bosonic representation. 

Finally, it is worth noting that, while not our primary focus, fermion-to-boson maps are themselves of interest for both practical and fundamental reasons \cite{Garbaczewski1978, Klein1991}. The current map achieves this in a simple form that exactly recovers the correct matrix structure of the many-fermion problem and avoids the issues in some previous ones that result in infinite expansions of fermion operators in terms of bosonic ones \cite{Garbaczewski1978, Klein1991}. We also note that other spin-to-boson maps are possible, such as the Holstein-Primakoff \cite{Holstein1940} and Matsubara-Matsuda \cite{Matsubara1956} transformations. In the Supplemental Material, Sec.~\ref{sec:alternative-spin-to-boson-maps}, we derive the Cartesian maps of fermionic operators that would be obtained using these transformations. We show that the former can also be used to obtain a phase space map that exactly recovers the matrix structure of the many-fermion problem, albeit at the price of cumbersome nonlinearities in the form of square roots of occupation number operators. For the latter, while we provide the Cartesian map that could be generated from it, we also show that this map is unable to yield an exact Cartesian representation of fermionic operators or their matrix elements. However, we suggest how it could be used in a controlled manner in a path integral treatment of many-fermion problems.

We are now in a position to map the fermionic operators $\{ \hat{c}_{j}, \hat{c}_{j}^{\dagger} \}$ to Cartesian phase space variables $\{\hat{q}_{j \gamma}, \hat{p}_{j \gamma}\}$. We achieve this by expressing the bosonic operators $\{ \hat{b}_{j \gamma}, \hat{b}_{j \gamma}^{\dagger} \}$ in Eqs.~(\ref{eq:JW-transformation-schwinger}) and (\ref{eq:definition-nonlocal-F-schwinger}) in their phase space representation,
\begin{subequations} \label{eq:bosons-as-cartesian}
    \begin{align}
    \hat{b} &= (\hat{q} + i\hat{p}) / \sqrt{2},\\
    \hat{b}^{\dagger} &= (\hat{q} - i\hat{p})/\sqrt{2}.
    \end{align}
\end{subequations}
This yields,
\begin{subequations} \label{eq:JW-transformation-cartesian}
    \begin{align}
    \hat{c}_j &\ ``\mapsto" \ \frac{1}{2}\hat{F}(0,j) (\hat{q}_{j\alpha} - i\hat{p}_{j\alpha})(\hat{q}_{j\beta} + i\hat{p}_{j\beta}), \\
    \hat{c}_j^{\dagger} &\ ``\mapsto"\ \frac{1}{2} \hat{F}(0,j) (\hat{q}_{j\beta} - i\hat{p}_{j\beta})(\hat{q}_{j\alpha} + i\hat{p}_{j\alpha}),
	\end{align}
\end{subequations}
where
\begin{equation}\label{eq:definition-nonlocal-F-cartesian}
    \begin{split}
    \hat{F}(j,k) &\ ``\mapsto" \ \prod_{l=\min[j+1,k+1]}^{\max[j-1,k-1]} f(\hat{m}_{l\beta} - \hat{m}_{l\alpha}),
    \end{split}
\end{equation}
and 
\begin{equation}\label{eq:bosonic-occupation-cartesian}
\hat{m}_{j\gamma} = \frac{1}{2}\Big(\hat{q}_{j \gamma}^2 + \hat{p}_{j\gamma}^2 - 1\Big)
\end{equation}
can be recognized as the occupation number operator corresponding to the $\gamma \in \{\alpha, \beta \}$ boson labelled by index $j$. Equations (\ref{eq:JW-transformation-cartesian}) and (\ref{eq:definition-nonlocal-F-cartesian}) then allow us to map $\hat{H}$ in Eq.~(\ref{eq:general-fermion-hamiltonian}) to a Cartesian representation, $\hat{H}(\{ \hat{c}_j^{\dagger}, \hat{c}_j \}) \mapsto \hat{H}(\{ \hat{q}_{j\gamma}, \hat{p}_{j\gamma} \})$, expressed in terms of the continuous coordinates and momenta of fictitious particles, 
    \begin{widetext}
    \begin{equation} \label{eq:general-fermion-hamiltonian-mapped}
    \begin{split}
        \hat{H} &= \frac{1}{2}\sum_{j,k} h_{j,k}(\boldsymbol{\Gamma}) \hat{F}(j,k)(\hat{q}_{j\beta} - i\hat{p}_{j\beta})(\hat{q}_{j\alpha} + i\hat{p}_{j\alpha})(\hat{q}_{k\beta} + i\hat{p}_{k\beta})(\hat{q}_{k\alpha} - i\hat{p}_{k\alpha}) \\
        &\quad + \frac{1}{2^3} \sum_{j,k,l,m} U_{jk,lm}(\boldsymbol{\Gamma}) \mathrm{sgn}(k-j)\mathrm{sgn}(m-l) \hat{F}(j,k,l,m) (\hat{q}_{j\beta} - i\hat{p}_{j\beta})(\hat{q}_{j\alpha} + i\hat{p}_{j\alpha}) (\hat{q}_{k\beta} - i\hat{p}_{k\beta})(\hat{q}_{k\alpha} + i\hat{p}_{k\alpha})\\
        &\qquad \qquad \qquad \qquad \qquad \qquad \qquad \qquad \qquad \qquad \qquad \qquad \times (\hat{q}_{m\beta} + i\hat{p}_{m\beta})(\hat{q}_{m\alpha} - i\hat{p}_{m\alpha}) (\hat{q}_{l\beta} + i\hat{p}_{l\beta})(\hat{q}_{l\alpha} - i\hat{p}_{l\alpha}),
    \end{split}
    \end{equation}
    \end{widetext}
where
\begin{equation}\label{eq:multi-index-nonlocal-F}
    \hat{F}(\mathbf{r}) \equiv \hat{F}(s_1, s_2) \hat{F}(s_3, s_4) \times ... \times \hat{F}(s_{2N-1}, s_{2N})
\end{equation}
is the many-index generalization of the antisymmetry operator, $\mathbf{s} = \{ s_1, ..., s_{2N} \}$ corresponds to the $2N$ members of $\mathbf{r} = \{ r_1, ..., r_{2N} \}$ arranged in increasing order, and $\mathrm{sgn}(x)$ returns the sign of its argument, $x$. As we show in the Supplemental Material, Sec.~\ref{sec:fermion-to-boson-map}, to obtain Eq.~(\ref{eq:general-fermion-hamiltonian-mapped}), one transforms the original many-fermion Hamiltonian to its representation in terms of bosonic creation and annihilation operators, places them in normal ordered form (where all creation operators lie to the left), and then truncates the unphysical excitations. Hence, by starting from a series of well-defined transformations, we have obtained a quantum mechanically exact representation of the fermionic matrix elements that is applicable to general many-fermion Hamiltonians.

The anticommutivity operator, $\hat{F}(j,k)$, clearly plays a vital role in our map (Eqs.~(\ref{eq:JW-transformation-cartesian})-(\ref{eq:definition-nonlocal-F-cartesian})) since it allows one to transform fermionic operators to bosonic ones while avoiding the exponential scaling associated with the mapping of the many-body basis. However, since the scaling reduction comes at the price of introducing this nonlocal operator, it is important to consider how its influence manifests in mapped operators and systems. For example, consider the evaluation of an arbitrary quadratic operator, $\hat{c}_j^{\dagger}\hat{c}_k$, in its mapped form $\hat{F}(j,k) \hat{b}_{j\beta}^{\dagger}\hat{b}_{j\alpha}\hat{b}_{k\alpha}^{\dagger}\hat{b}_{k\beta}$. In its mapped form, one can exploit the factorization of the many-body basis of bosons which allows for the calculation of the matrix elements of $\hat{F}(j,k)\hat{b}_{j\beta}^{\dagger}\hat{b}_{j\alpha}\hat{b}_{k\alpha}^{\dagger}\hat{b}_{k\beta}$ with $2|j-k|+2$ single-body inner products: 4 boson modes corresponding to the 2 indices $j$ and $k$ and the product of $2|j-k|-2$ single-body bosonic operators in $\hat{F}(j,k)$. In contrast, if $\hat{F}(j,k)$ were absent, the matrix elements of the mapped operator would require only $4$ single-body inner products. Hence, minimizing the presence of $\hat{F}(j,k)$ in mapped operators and Hamiltonians is advantageous. Below, we demonstrate how one can exploit the choice of index ordering in the JW transformation to curb the nonlocality of $\hat{F}$ and in some cases eliminate it completely. To illustrate this, we show how our map can be applied to the Anderson and Hubbard models, which are representative of fermionic systems belonging to the impurity and lattice families, respectively.

We start with the Anderson model,
\begin{equation}\label{eq:anderson}
    \begin{split}
    \hat{H}_{\mathrm{And}} &= \sum_{\lambda} \varepsilon_{\lambda} \hat{\tilde{n}}_{\lambda} + \sum_{u, \lambda, a} \varepsilon_{u, \lambda, a} \hat{n}_{u, \lambda, a} + U \hat{\tilde{n}}_{\uparrow} \hat{\tilde{n}}_{\downarrow} \\
    &\quad  + \sum_{u, \lambda, a} t_{u, \lambda, a} \big[ \hat{c}_{u, \lambda, a}^{\dagger} \hat{\tilde{c}}_{\lambda} + \hat{\tilde{c}}_{\lambda}^{\dagger}\hat{c}_{u, \lambda, a}\big],
    \end{split}
\end{equation}
where an impurity, which can accommodate an interacting pair of spin up and down electrons, is coupled to two leads per spin at (possibly) different temperatures and/or chemical potentials. Here, operators with an additional tilde $\{ \hat{\tilde{c}}, \hat{\tilde{c}}^{\dagger}, \hat{\tilde{n}} \}$ correspond to the impurity, while all others correspond to the fermions in the leads, $\hat{n}_{u} = \hat{c}_{u}^{\dagger}\hat{c}_u$ is the occupation number operator, $u$ labels the single-particle orbitals that comprise the leads, $\lambda \in \{\uparrow, \downarrow \}$ labels the spin, and $a \in \{\mathrm{R}, \mathrm{L} \}$ distinguishes the right ($R$) from left ($L$) spin-dependent leads. The parameters of the Hamiltonian include the single-electron terms consisting of the impurity and lead state energies $\varepsilon_{\lambda}$ and $\varepsilon_{j, \lambda}$ and the coupling (impurity-lead hybridization) terms, $t_{j, \lambda}$, connecting the impurity and lead states, while $U$ is the two-body Coulomb term. 

\begin{figure}
    \centering
    \includegraphics[width=\columnwidth]{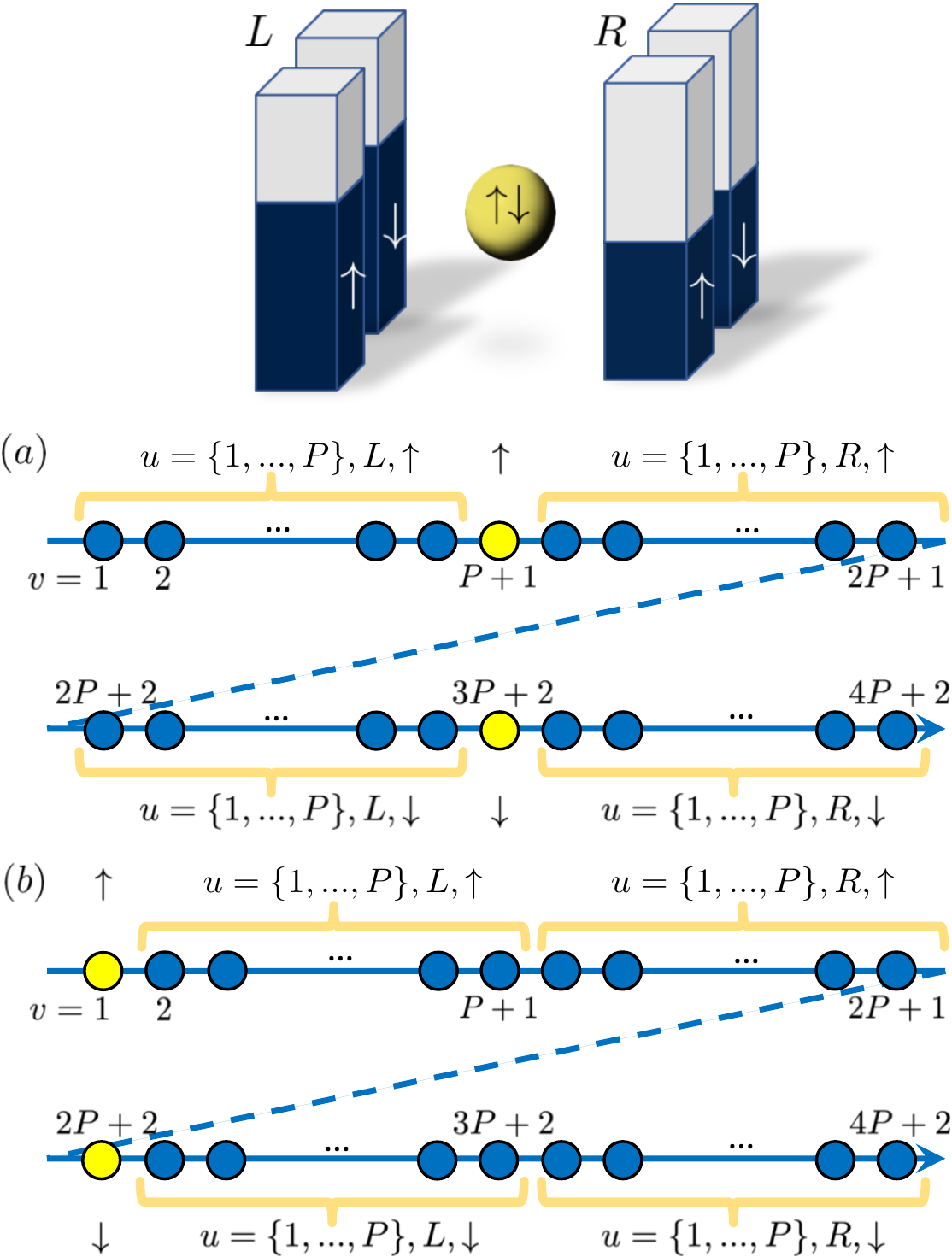}
    \caption{Index ordering options for the Anderson model.} 
    \label{fig:anderson}
\end{figure}

Using Eqs.~(\ref{eq:JW-transformation-cartesian})-(\ref{eq:definition-nonlocal-F-cartesian}), one can map the Anderson Hamiltonian, Eq.~(\ref{eq:anderson}), as
\begin{equation}\label{eq:mapped-anderson}
    \begin{split}
    \hat{H}_{\mathrm{And}} &\mapsto \sum_{\lambda} \varepsilon_{\lambda} \hat{\tilde{\eta}}_{\lambda} + \sum_{u, \lambda, a} \varepsilon_{u, \lambda, a} \hat{\eta}_{u, \lambda, a} + U \hat{\tilde{\eta}}_{\uparrow} \hat{\tilde{\eta}}_{\downarrow}\\
    & + \frac{1}{2}\sum_{u, \lambda, a} \hat{F}_{\lambda, a}(0,u) t_{u, \lambda, a}  \big[ \hat{\tilde{x}}_{\lambda}\hat{x}_{u,\lambda,a} + \hat{\tilde{y}}_{\lambda}\hat{y}_{u,\lambda,a} \big],
    \end{split}
\end{equation}
    where
\begin{subequations} \label{eq:mapped-auxiliary-operators}
    \begin{align}
    \hat{\eta}_{u} &\equiv \hat{m}_{u, \beta} \big(\hat{m}_{u, \alpha} + 1\big), \label{eq:mapped-occupation-number}\\
    \hat{x}_{u} &\equiv \big(\hat{q}_{u\beta}\hat{q}_{u\alpha} + \hat{p}_{u\beta}\hat{p}_{u\alpha}\big), \label{eq:mapped-x}\\
    \hat{y}_{u} &\equiv \big(\hat{q}_{u\beta}\hat{p}_{u\alpha} - \hat{q}_{u\alpha}\hat{p}_{u\beta}\big), \label{eq:mapped-y}
    \end{align}
\end{subequations}
and $\hat{m}_{j}$ is defined in Eq.~(\ref{eq:bosonic-occupation-cartesian}). We emphasize that, since every fermionic creation and annihilation operator is mapped onto two coupled oscillators, quadratic operators, such as the occupation number operator, $\hat{n}_{u} \mapsto \hat{\eta}_{u}$, result in expressions that are quadratic in the oscillator occupation operators and quartic in the phase space variables, i.e., $\hat{m}_{u, \beta}(\hat{m}_{u, \alpha}+1) = (\hat{q}_{u, \beta}^2 + \hat{p}_{u, \beta}^2 - 1)(\hat{q}_{u, \alpha}^2 + \hat{p}_{u, \alpha}^2 + 1)/ 4$. By using the Matsubara-Matsuda transformation to map the spins in Eq.~(\ref{eq:sigmas-schwinger}), one can construct a phase space representation that yields only quadratic terms in the phase space variables. However, as we show in the Supplemental Material, Sec.~\ref{ssec:matsubara-matsuda}, such a map would not be exact. We also note that only the final term in Eq.~(\ref{eq:mapped-anderson}) contains the nonlocal operator $\hat{F}$, since, for occupation number operators, the mapped product of creation and annihilation operators on the same site trivially removes the nonlocal component, regardless of ordering.

To obtain the mapped version of any Hamiltonian, one must perform the JW transformation, which maps all the single-particle states that constitute the system to a 1D ordered spin chain i.e., with a single index. Hence, in the context of the Anderson model, it is necessary to choose an index ordering that both collapses the labels over the single-particle identifier, fermion spin, and lead label onto a single index, and minimizes the extent to which the nonlinear operator $\hat{F}$ appears in the mapped Hamiltonian. To achieve this, we arrange the single-particle orbitals along a chain corresponding first to $\lambda = \uparrow$, starting with the $a = R$ lead states, continuing with the impurity state, and ending with the $a = L$ lead states. A similar arrangement is then chosen for the $\lambda = \downarrow$ states. In principle, the leads correspond to infinitely large sources of electrons. However, in practice, one discretizes them into a suitably large number, $P$, of single-particle orbitals. This results in the 1D indexing: for the lead states $v = u + \delta_{\lambda,\uparrow} \delta_{a,R} (P+1) + \delta_{\lambda,\downarrow} [\delta_{a,R} (2P+1) +  \delta_{a,L} (3P+2)]$ and for the impurity orbitals $v = \delta_{\lambda, \uparrow} (P+1) + \delta_{\lambda, \downarrow} (3P+2)$. This indexing is depicted in Fig.~\ref{fig:anderson}a.

Because there are no spin-flip terms (e.g., $\hat{c}^{\dagger}_{u, \lambda, a}\hat{c}_{u\prime, \lambda\prime, a\prime}$) in this Hamiltonian, this indexing allows one to separate the spin chain into two parts which are not connected by the nonlocal operator $\hat{F}(v, v^{\prime})$. This ability to separate the spin chain means that $\hat{F}(v, v^{\prime})$ can be written as simply $\hat{F}_{\uparrow, a}(0, u)$ or $\hat{F}_{\downarrow, a}(0, u)$ (i.e., containing only connections within a particular lead) without introducing cross terms.

To illustrate the importance of the choice of index ordering, one can consider the effect of instead starting with $\lambda = \uparrow$, going from the impurity to the $a = R$ and $a = L$ orbitals, and then continuing with the $\lambda = \downarrow$ orbitals in the same fashion (Fig.~\ref{fig:anderson}b). While this choice of index ordering still provides a formally exact mapping, it results in the hopping terms connecting the impurity and left leads being modified by the occupations in the right leads, i.e., $t_{u,\lambda, L}\hat{F}_{\lambda, R}(0, P)\hat{F}_{\lambda, L}(0, u) $. Hence, it is important to consider which ordering leads to the simplest version of the mapped operators. 

We now turn to the 1D Hubbard model,
\begin{equation}\label{eq:hubbard}
    \begin{split}
    \hat{H}_{\mathrm{Hub}} &= \sum_{u} U_{u} \hat{n}_{\uparrow,u}\hat{n}_{\downarrow,u} \\
    &\qquad + \sum_{u, \lambda} t^{(\lambda)}_{u,u+1} \big[\hat{c}_{\lambda, u}^{\dagger}\hat{c}_{\lambda, u+1} + \hat{c}_{\lambda, u+1}^{\dagger}\hat{c}_{\lambda, u}\big],
    \end{split}
\end{equation}
which consists of a chain of sites that can accommodate interacting spin up and down fermions with nearest neighbor coupling. Here, $\lambda \in \{ \uparrow, \downarrow \}$ is the spin index, $u$ is the spatial index of the sites along the Hubbard chain, $t_{u,u+1}^{(\lambda)}$ is the one-electron nearest neighbor hopping term, and $U_{u}$ is the two-body Coulomb repulsion term, analogous to $h_{j,k}$ and $U_{jk,lm}$ in Eq.~(\ref{eq:general-fermion-hamiltonian}), respectively.

Using Eqs.~(\ref{eq:JW-transformation-cartesian})-(\ref{eq:definition-nonlocal-F-cartesian}), one can map the Hubbard Hamiltonian, Eq.~(\ref{eq:hubbard}), as
\begin{equation} \label{eq:mapped-hubbard}
    \begin{split}
    \hat{H}_{\mathrm{Hub}} &\mapsto \sum_{u} U_{u} \hat{\eta}_{\uparrow,u}\hat{\eta}_{\downarrow,u} \\
    &\quad + \frac{1}{2}\sum_{u, \lambda} t^{(\lambda)}_{u,u+1} \big[\hat{x}_{\lambda,u}\hat{x}_{\lambda,u+1} + \hat{y}_{\lambda,u}\hat{y}_{\lambda,u+1}\big],
    \end{split}
\end{equation}
where $\hat{\eta}_{\lambda, u}$, $\hat{x}_{\lambda, u}$ and $\hat{y}_{\lambda, u}$ are defined in Eqs.~(\ref{eq:mapped-occupation-number})-(\ref{eq:mapped-y}). In practice, the infinite Hubbard chain is truncated to $P$ sites. 

As for the Anderson model, the first term, which contains only occupation number operators, does not have nonlocal contributions, $\hat{F}$, regardless of the index ordering, while the hopping term generally contains them. However, one can remove them with a judicious choice of index ordering. To do this, we choose the fermion indices as $v = u + \delta_{\lambda, \uparrow} P$, where $u \in \{ 1, 2, ..., P\}$ labels the site number along the Hubbard chain. Following this indexing prescription, shown in Fig.~\ref{fig:hubbard}a, one can exploit the nearest neighbor coupling in the 1D Hubbard model to obtain a mapped Hamiltonian that is completely free from the influence of the nonlocal operator $\hat{F}$. As for the Anderson model, we then transform from $v$ to the original indices $u$ and $\lambda$, leading to Eq.~(\ref{eq:mapped-hubbard}). 

\begin{figure}
    \centering
    \includegraphics[width=\columnwidth]{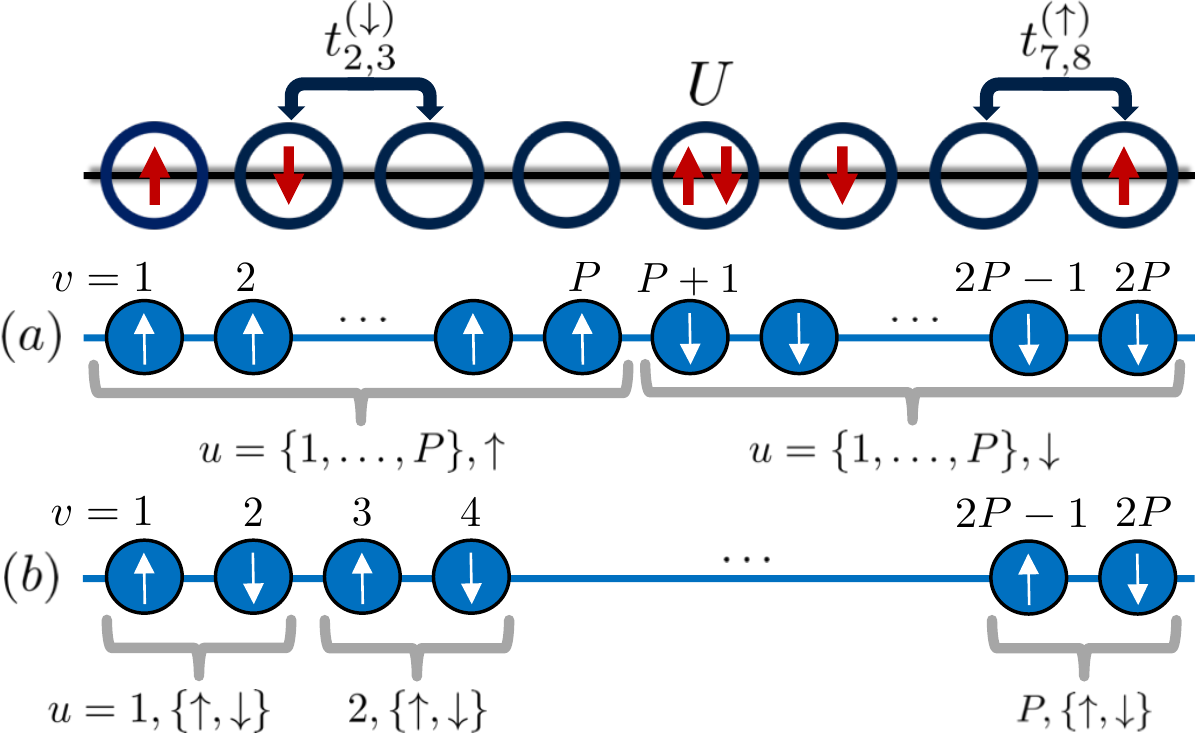}
    \caption{Index ordering options for the 1D Hubbard model.} 
    \label{fig:hubbard}
\end{figure}

In contrast, if one were to order the chain according to the scheme shown in Fig.~\ref{fig:hubbard}b, intercalating the up and down spins as one sweeps from left to right across the chain, i.e., $v = (2u-1)\delta_{\lambda, \uparrow} + 2u\delta_{\lambda, \downarrow}$, one would not be able to eliminate all the nonlocal terms. Indeed, this choice would result in a renormalization of the hopping terms by an operator that tracks the occupation of one of the two neighboring sites with a fermion of the opposite spin, i.e., $t_{u, u+1}^{(\uparrow)} \mapsto t_{u, u+1}^{(\uparrow)} f(\hat{m}_{\downarrow,u,\beta} - \hat{m}_{\downarrow,u,\alpha})$ and  $t_{u, u+1}^{(\downarrow)} \mapsto t_{u, u+1}^{(\downarrow)} f(\hat{m}_{\downarrow,u+1,\beta} - \hat{m}_{\downarrow,u+1,\alpha})$. It is thus noteworthy that when the transformation introduced here is employed with a judicious choice of indexing, one can rewrite the 1D Hubbard model entirely as a continuous bosonic Hamiltonian devoid of the nonlocal operator $\hat{F}$, which is consequently free from the complexities arising from the sign flips associated with fermionic anticommutivity.

In conclusion, we have derived a map that expresses fermionic creation and annihilation operators in terms of continuous, Cartesian phase space variables. Importantly, this representation captures the exact matrix structure of the many-fermion problem. We have then shown how one can apply the map introduced here to the Anderson and 1D Hubbard models in a way that minimizes, and in some cases completely eliminates, the nonlocal fermionic anticommutivity operator, $\hat{F}$. 

Finally, it is worth contrasting this work with previous approaches that can in principle exactly describe many-fermion problems in continuous phase space. These fall into two distinct categories. The first is based on bosonic coherent states and exploits the variational principle to describe the dynamics of a many-body antisymmetrized wavefunction ansatz \cite{Ono1992, Feldmeier2000, Kirrander2011}, at the price of poor scaling with system size \cite{Kirrander2011}. The second is based on fermionic coherent states and focuses on second-quantized operators instead of the wavefunction \cite{Plimak2001, Corney2004, Corney2006, Corney2006a, Dalton2016, Polyakov2016}. However, when considering individual fermionic creation or annihilation operators, it becomes necessary to use Grassmann variables, which are objects of high computational complexity whose phase space distributions can present interpretational difficulties that require subsequent mapping to a complex number phase space \cite{Corney2006, Polyakov2016, Davidson2017}. In contrast, our approach is fully compatible with bosonic and spin coherent states \cite{Klauder1997, coherent-states-book}, which circumvents the difficulties posed by Grassmann variables while also benefiting from the improved scaling that arises from mapping the individual fermionic creation and annihilation operators rather than the many-body wavefunctions.

The \emph{exact} Cartesian representation of fermionic creation and annihilation operators introduced here provides a starting point to employ methods that exploit classical-like trajectories to calculate static and dynamic quantum properties of many-fermion systems, including exact path integrals and approximate semiclassical and quantum-classical theories. This mapping enables the systematic development of a hierarchy of trajectory-based quantum dynamical methods of varying cost and accuracy that can be tuned to the scale and requirements of the physical problem under consideration. By rendering the many-fermion problem in a form that is compatible with broadly applicable trajectory-based methods, our approach thus provides an avenue for the study of many problems that are currently inaccessible to existing methods, including those where fermionic degrees of freedom are coupled to complex and anharmonic nuclear motion and spins.  

\begin{acknowledgments}
This material is based upon work supported by the U.S. Department of Energy, Office of Science, Office of Basic Energy Sciences under Award Number DE-SC0014437. T.E.M also acknowledges support from a Cottrell Scholarship from the Research Corporation for Science Advancement and the Camille Dreyfus Teacher-Scholar Awards Program.
\end{acknowledgments}

%

%% Supplemental Material

\clearpage

\setcounter{section}{0}
\setcounter{equation}{0}
\setcounter{figure}{0}
\setcounter{table}{0}
\setcounter{page}{1}

\renewcommand{\theequation}{S\arabic{equation}}
\renewcommand{\thefigure}{S\arabic{figure}}
\renewcommand{\thepage}{S\arabic{page}}
\renewcommand{\bibnumfmt}[1]{$^{\mathrm{S#1}}$}
\renewcommand{\citenumfont}[1]{S#1}

\title{Supplemental Material: On the exact continuous mapping of fermions}
{\maketitle}

\onecolumngrid

%\clearpage
%\onecolumngrid
%
%\section{Supplemental Material: On the exact continuous mapping of fermions} 
%
%\appendix
%
%\setcounter{figure}{0}
%\makeatletter 
%\renewcommand{\thefigure}{S\@arabic\c@figure}
%\makeatother
%\renewcommand{\theequation}{S\arabic{equation}}

In Sec. \ref{sec:schwinger-theory-basics}, we briefly review the necessary background of the Schwinger theory of angular momentum and demonstrate how it is exact at the matrix element level but not the operator level. In Sec.~\ref{sec:fermion-to-boson-map}, we provide a derivation of how one can use the JW transformation and the Schwinger theory of angular momentum to obtain a fermion-to-boson map that exactly reproduces the matrix structure of the many-fermion problem. Importantly, in this section, we demonstrate that all the spurious excitations that arise as a result of the Schwinger mapping are eliminated when the physical basis is used to evaluate the matrix elements, which ensures that \emph{both} the forward and backward fermion-to-spin-to-fermion transformations are well defined. In Sec.~\ref{sec:alternative-spin-to-boson-maps}, we show that, for the purpose of obtaining an exact and practical Cartesian mapping of fermionic operators, two alternative spin-to-boson transformations that one may consider suffer from serious deficiencies.

\section{Schwinger theory of angular momentum: spin-boson isomorphism at the matrix element level}
\label{sec:schwinger-theory-basics}
	
In this section we demonstrate and stress how the Schwinger theory of angular momentum provides an isomorphic representation of the spin algebra at the matrix element level, but \emph{not} at the operator level. The Schwinger boson representation \cite{Schwinger1965-sup} expresses spin $1/2$ operators in terms of two coupled bosons per spin degree of freedom,
\begin{subequations} \label{eq:sigmas-schwinger-si}
    \begin{eqnarray}
    \sigma_j^+ = \ket{1_j}\bra{0_j} \quad &``\xmapsto[]{Schwinger} "& \quad \hat{b}^{\dagger}_{j\beta}\hat{b}_{j\alpha}, \label{eq:sigma+schwinger-si} \\
    \sigma_j^- = \ket{0_j}\bra{1_j} \quad &``\xmapsto[]{Schwinger} "& \quad \hat{b}^{\dagger}_{j\alpha}\hat{b}_{j\beta}, \label{eq:sigma-schwinger-si} \\
    \mathbf{1}_j = \ket{1_j}\bra{1_j} + \ket{0_j}\bra{0_j} \quad &``\xmapsto[]{Schwinger} "& \quad \hat{b}^{\dagger}_{j\beta}\hat{b}_{j\beta} + \hat{b}^{\dagger}_{j\alpha}\hat{b}_{j\alpha} \label{eq:spin-unit-schwinger-si}\\
    \sigma_j^{z} = \ket{1_j}\bra{1_j} - \ket{0_j}\bra{0_j} \quad &``\xmapsto[]{Schwinger} "& \quad  \hat{b}^{\dagger}_{j\beta}\hat{b}_{j\beta} - \hat{b}^{\dagger}_{j\alpha}\hat{b}_{j\alpha}, \label{eq:sigmaz-schwinger-si}
    \end{eqnarray}
\end{subequations}
where the coupled bosons follow regular bosonic commutation relations,
\begin{subequations} \label{eq:boson-commutation-relations}
	\begin{eqnarray}
	[\hat{b}_{j\alpha}, \hat{b}_{k\beta}^{\dagger}] &=& \delta_{j,k}\delta_{\alpha , \beta}, \label{eq:bosonic-commutation-relations-mixed} \\ 
	\ [\hat{b}_{j\alpha}, \hat{b}_{k\beta}] &=& 0. \label{eq:bosonic-commutation-relations-same}
	\end{eqnarray}
\end{subequations}
The quotation marks around the map symbol in Eqs.~(\ref{eq:sigmas-schwinger-si}) connecting the spin operators and the coupled bosons are used to stress the fact that, on the operator level, this transformation only recovers the correct result when double and higher-order (joint and individual) excitations of the $\alpha$ and $\beta$ modes corresponding to a given spin index $j$ are excluded. To demonstrate this point, consider the commutation and anticommutation relations of spin operators: similar to bosons, spins on different sites commute, 
\begin{equation}\label{eq:spin-commutation}
	[\tilde{\sigma}_j, \tilde{\sigma}_k] = 0,
\end{equation}
where $\tilde{\sigma}_j \in \{\sigma_j^+, \sigma_j^-, \sigma_j^z \}$ and $j \neq k$; in contrast, unlike bosons, spins on the same site obey the following commutation and anticommuation relations,
\begin{subequations}\label{eq:same-site-spin-commutation-relations}
	\begin{eqnarray}
	[\sigma_j^+, \sigma_j^-] &=& \sigma_j^z, \label{eq:spin-commutation-ladders} \\
	\ [\sigma_j^z, \sigma_j^{\pm}] &=& \pm 2\sigma_j^{\pm},\label{eq:spin-commutation-ladder-sz}
	\end{eqnarray}
\end{subequations}
and
\begin{equation}\label{eq:spin-anticommutation}
	\{ \sigma_j^+, \sigma_j^- \} = \mathbf{1}_j.
\end{equation}
Using the Schwinger boson representation in Eq.~(\ref{eq:sigmas-schwinger-si}), one can easily recover the spin commutation relations, Eqs.~(\ref{eq:spin-commutation}) and (\ref{eq:same-site-spin-commutation-relations}). However, the anticommutation relation in Eq.~(\ref{eq:spin-anticommutation}) is only approximately recovered, 
\begin{equation}\label{eq:spin-anticommutation-relations-schwinger-bosons}
	\{ \sigma_j^+, \sigma_j^- \} \quad ``\xmapsto[]{Schwinger} " \quad \hat{b}^{\dagger}_{j\beta}\hat{b}_{j\beta} + \hat{b}^{\dagger}_{j\alpha}\hat{b}_{j\alpha} + 2 \hat{b}^{\dagger}_{j\beta}\hat{b}^{\dagger}_{j\alpha}\hat{b}_{j\alpha}\hat{b}_{j\beta} = \mathbf{1}_j + \mathcal{O}_j(2),
\end{equation}
where, $\mathcal{O}_j(n)$ denotes terms containing $n^{\mathrm{th}}$ order excitations for the $j^{\mathrm{th}}$ index. This result is obtained by placing the boson creation and annihilation operators in a \emph{normal-ordered} form, i.e., where the creation operators lie to the left. These high-order excitations could be joint $\alpha$ and $\beta$ mode excitations, such as the double excitation $\hat{b}_{j\beta}^{\dagger}\hat{b}_{j\alpha}^{\dagger}\hat{b}_{j\alpha}\hat{b}_{j\beta}$, or individual mode excitations, such as the second order excitation $\hat{b}_{j\gamma}^{\dagger}\hat{b}_{j\gamma}^{\dagger}\hat{b}_{j\gamma}\hat{b}_{j\gamma}$, where $\gamma \in \{ \alpha, \beta \} $. Hence, Eq.~(\ref{eq:spin-anticommutation-relations-schwinger-bosons}) shows that the Schwinger boson representation only recovers the algebra of spin operators on the joint single-excitation subspace for the $j^{\mathrm{th}}$ bosonic modes (the subspace for the $j^{\mathrm{th}}$ coupled bosons consisting of the states where the $\alpha$ mode is in the ground state while the $\beta$ mode is in the first excited state, and vice versa). Therefore, \emph{the Schwinger boson representation for spins only represents an exact isomorphism at the level of matrix elements evaluated using the ``physical'' single-excitation basis of the bosons corresponding to the $j$ spin.}

\section{Mapping Fermions to Schwinger-bosons: Proof of Exactness at the Matrix Element Level}
\label{sec:fermion-to-boson-map}

In this section, we show how one can map fermionic creation and annihilation operators to bosonic ones, and thereby Cartesian phase space variables, by sequentially exploiting the JW transformation and the Schwinger theory of angular momentum, and demonstrate that the resulting map constitutes an exact isomorphism that reproduces the matrix structure of the many-fermion problem.

Using the Schwinger representation in Eq.~(\ref{eq:sigmas-schwinger-si}) to express the JW transformed fermionic creation and annihilation operators in Eqs.~(\ref{eq:JW-transformation})-(\ref{eq:definition-nonlocal-F}), we map fermionic operators to boson and Cartesian phase space operators,
\begin{subequations} \label{eq:JW-transformation-schwinger-si}
    \begin{align}
    	\hat{c}_j  \quad  &``\mapsto " \quad  \hat{F}(0,j) \hat{b}^{\dagger}_{j\alpha}\hat{b}_{j\beta} = \frac{1}{2}\hat{F}(0,j) (\hat{q}_{j\alpha} - i\hat{p}_{j\alpha})(\hat{q}_{j\beta} + i\hat{p}_{j\beta}), \label{eq:JW-transformation-annihilation-schwinger-si} \\
    	\hat{c}_j^{\dagger}  \quad &``\mapsto " \quad  \hat{F}(0,j) \hat{b}^{\dagger}_{j\beta}\hat{b}_{j\alpha} = \frac{1}{2} \hat{F}(0,j) (\hat{q}_{j\beta} - i\hat{p}_{j\beta})(\hat{q}_{j\alpha} + i\hat{p}_{j\alpha}), \label{eq:JW-transformation-creation-schwinger-si}
	\end{align}
\end{subequations}
where
    \begin{equation}
        \hat{F}(j,k)  \quad ``\mapsto " \quad  \prod_{l=\min[j+1,k+1]}^{\max[j-1,k-1]} f(\hat{b}^{\dagger}_{j\beta}\hat{b}_{j\beta} - \hat{b}^{\dagger}_{j\alpha}\hat{b}_{j\alpha}) =  \prod_{l=\min[j+1,k+1]}^{\max[j-1,k-1]} f\Bigg(\frac{1}{2}\big[\hat{q}_{l\alpha}^2 + \hat{p}_{l\alpha}^2 - \hat{q}_{l\beta}^2 - \hat{p}_{l\beta}^2\big]\Bigg). \label{eq:definition-nonlocal-F-schwinger-si}
    \end{equation}

As in case of the Schwinger transformation, we keep the quotation marks around the map symbol connecting the fermionic operators and the bosonic representation in Eqs.~(\ref{eq:JW-transformation-schwinger-si}) and (\ref{eq:definition-nonlocal-F-schwinger-si}) to indicate that this transformation can be expected to recover the correct fermionic algebra on the joint single excitation subspace of the coupled $\alpha$ and $\beta$ bosons per fermionic index $j$. For more complex operators, $\hat{O}(\{ \hat{c}_j^{\dagger}, \hat{c}_j \})$, constructed from products of fermionic creation and annihilation operators, the Schwinger mapped version will in general contain such higher order excitations, 
\begin{equation} \label{eq:arbitrary-operator-map-schwinger}
    \hat{O}(\{ \hat{c}_j^{\dagger}, \hat{c}_j \}) \quad \xmapsto[]{JW} \quad \hat{O}(\{ \tilde{\sigma}_j \})  \quad ``\xmapsto[]{Schwinger} " \quad  \hat{O}(\{ \hat{b}_{j\beta}^{\dagger}, \hat{b}_{j\beta}, \hat{b}_{j\alpha}^{\dagger}, \hat{b}_{j\alpha} \}). 
    \end{equation}
Equations (\ref{eq:JW-transformation-schwinger-si})-(\ref{eq:arbitrary-operator-map-schwinger}) thus provide the procedure to start from an arbitrary operator composed of fermionic creation and annihilation operators, transition via the JW transformation to an equivalent expression in terms of spin operators, and then use the Schwinger representation to render the JW-transformed fermionic operator in a bosonic form (which can, in turn, be mapped to a Cartesian representation as shown in the main paper). 

In the following subsections, we show that the final step of the fermion-to-spin-to-boson transformation in Eq.~(\ref{eq:arbitrary-operator-map-schwinger}) does not introduce spurious terms that would prevent the inverse boson-to-spin-to-fermion transformation. To do this, we demonstrate that the deviations from the exact fermionic operator-level map that arise due to the Schwinger representation occur exclusively in the unphysical space of high-lying excitations and are naturally excluded when using the physical basis. 

\subsection{Emergence of unphysical excitations in the fermion-to-boson map}
\label{ssec:unphysical-excitations}

Here, we demonstrate that arbitrary products of spin operators in the Schwinger representation \emph{always} lead to the correct mapped spin operators on the single-excitation manifold but that these are generally accompanied with spurious additional high-order excitations. In this subsection our goal is not to eliminate these high-lying excitations (which will be done in Sec.~\ref{ssec:removal-of-excitations} by using the physical basis), but rather to show that the Schwinger representation always leads to the correct spin behavior on the single-excitation manifold and only results in deviations in subspaces corresponding to higher order excitations. 
Essential to this is that, in the Schwinger representation, unphysical excitations are never able to return to the single-excitation manifold. One can understand this by observing that the Schwinger representation of the basic spin operators (ladder operators), $\sigma_j^{+} \mapsto \hat{b}_{j\beta}^{\dagger}\hat{b}_{j\alpha}$ and $\sigma_j^-\mapsto \hat{b}_{j\alpha}^{\dagger}\hat{b}_{j\beta}$, requires that as the $\alpha$ mode is excited, the $\beta$ mode must be de-excited, and vice versa.

To illustrate this point in more detail, one can start by recognizing that the powers $n \geq 2$ of the mapped operators in Eqs.~(\ref{eq:sigmas-schwinger-si}), take the form
\begin{subequations} \label{eq:sigmas-schwinger-powers}
    \begin{eqnarray}
    [\sigma_j^+]^n \quad &``\xmapsto[]{Schwinger} "& \quad [\hat{b}^{\dagger}_{j\beta}\hat{b}_{j\alpha}]^n = 0 + \mathcal{O}_j(n), \label{eq:sigma+schwinger-powers} \\
    \ [\sigma_j^-]^n  \quad &``\xmapsto[]{Schwinger} "& \quad [\hat{b}^{\dagger}_{j\alpha}\hat{b}_{j\beta}]^n = 0 + \mathcal{O}_j(n), \label{eq:sigma-schwinger-powers} \\
    \ [\mathbf{1}_j]^n  \quad &``\xmapsto[]{Schwinger} "& \quad [\hat{b}^{\dagger}_{j\beta}\hat{b}_{j\beta} + \hat{b}^{\dagger}_{j\alpha}\hat{b}_{j\alpha}]^n = \mathbf{1}_j + \mathcal{O}(2) + ... + \mathcal{O}(n), \label{eq:spin-unit-schwinger-powers}\\
    \ [\sigma_j^{z}]^{n} \quad &``\xmapsto[]{Schwinger} "& \quad [\hat{b}^{\dagger}_{j\beta}\hat{b}_{j\beta} - \hat{b}^{\dagger}_{j\alpha}\hat{b}_{j\alpha}]^{n} = \begin{cases}
    \mathbf{1}_j + \mathcal{O}(2) + ... + \mathcal{O}(n) & \mathrm{for}\ n\ \mathrm{even} \\
    \sigma^z_j + \mathcal{O}(2) + ... + \mathcal{O}(n) & \mathrm{for}\ n\ \mathrm{odd}
    \end{cases}.\label{eq:sigmaz-schwinger-powers} 
    \end{eqnarray}
\end{subequations}
As Eqs.~(\ref{eq:sigma+schwinger-powers})-(\ref{eq:sigmaz-schwinger-powers}) show, powers of the spin operators lead to the correct contribution in the single-excitation manifold with additional unphysical high-order excitations. Similarly, products of arbitrary powers of the Schwinger-mapped spin polarization and the spin ladder operators also recover the correct result on the single-excitation manifold accompanied by additional unphysical higher-order excitations,
\begin{subequations} \label{eq:spin-interactions-for-anticommutivity-even}
	\begin{eqnarray}
	[-\sigma_j^z]^n \sigma_j^{+}  \quad &``\xmapsto[]{Schwinger} "& \quad [\hat{b}^{\dagger}_{j\alpha}\hat{b}_{j\alpha} - \hat{b}^{\dagger}_{j\beta}\hat{b}_{j\beta} ]^{n} \hat{b}_{j\beta}^{\dagger}\hat{b}_{j\alpha} = (-1)^{n}\sigma_j^{+} + \mathcal{O}_j(2) + ... + \mathcal{O}_j(n), \label{eq:schwinger--sz-s+} \\
	\sigma_j^{+} [-\sigma_j^z]^n  \quad &``\xmapsto[]{Schwinger} "& \quad \hat{b}_{j\beta}^{\dagger}\hat{b}_{j\alpha}[\hat{b}^{\dagger}_{j\alpha}\hat{b}_{j\alpha} - \hat{b}^{\dagger}_{j\beta}\hat{b}_{j\beta}]^{n} = \sigma_j^{+} +  \mathcal{O}_j(2) + ... + \mathcal{O}_j(n), \label{eq:schwinger--s+-sz} \\
	\ [-\sigma_j^z]^n   \sigma_j^{-} \quad &``\xmapsto[]{Schwinger} "& \quad  [\hat{b}^{\dagger}_{j\alpha}\hat{b}_{j\alpha} - \hat{b}^{\dagger}_{j\beta}\hat{b}_{j\beta}]^{n} \hat{b}_{j\alpha}^{\dagger}\hat{b}_{j\beta} = \sigma_j^{-} + \mathcal{O}_j(2) + ... + \mathcal{O}_j(n) , \label{eq:schwinger--sz-s-} \\
	 \sigma_j^{-} [-\sigma_j^z]^n \quad &``\xmapsto[]{Schwinger} "& \quad \hat{b}_{j\alpha}^{\dagger}\hat{b}_{j\beta}[\hat{b}^{\dagger}_{j\alpha}\hat{b}_{j\alpha} - \hat{b}^{\dagger}_{j\beta}\hat{b}_{j\beta}]^{n} =  (-1)^{n}\sigma_j^{-} + \mathcal{O}_j(2) + ... + \mathcal{O}_j(n), \label{eq:schwinger--s--sz}  
	\end{eqnarray}
\end{subequations}
Equations~(\ref{eq:sigmas-schwinger-powers}) and (\ref{eq:spin-interactions-for-anticommutivity-even}) further demonstrate that the Schwinger representation correctly captures the spin algebra on the single-excitation manifold and that no operation arising from any product of mapped spin operators can cause an unphysical excitation to return to the physical single-excitation subspace. In other words, the product of an unphysical excitation of order $n$, $\mathcal{O}_j(n)$ of the type appearing in Eqs.~(\ref{eq:sigmas-schwinger-powers})-(\ref{eq:spin-interactions-for-anticommutivity-even}) and any mapped spin operator $\tilde{\sigma}_j \in \{ \sigma_j^+, \sigma_j^-, \sigma_j^z, \mathbf{1}_j \}$ \emph{always} leads to an excitation of similar order or higher, 
\begin{equation} \label{eq:higher-order-excitations-never-come-back}
	\begin{split}
	\tilde{\sigma}_j(\hat{b}_{j\beta}^{\dagger}, \hat{b}_{j\beta}, \hat{b}_{j\alpha}^{\dagger}, \hat{b}_{j\alpha}) \mathcal{O}_j(n) &= \mathcal{O}_j(n) + \mathcal{O}_j(n+1),\\
	\mathcal{O}_j(n)\tilde{\sigma}_j(\hat{b}_{j\beta}^{\dagger}, \hat{b}_{j\beta}, \hat{b}_{j\alpha}^{\dagger}, \hat{b}_{j\alpha}) &= \mathcal{O}_j(n) + \mathcal{O}_j(n+1).
	\end{split}
\end{equation}
Thus, as stated above, an unphysical high-order excitation ($n \geq 2$) can \emph{never} return to the physical single-excitation subspace.

Equations (\ref{eq:sigmas-schwinger-powers}), (\ref{eq:spin-interactions-for-anticommutivity-even}), and (\ref{eq:higher-order-excitations-never-come-back}) thus establish that the correct spin structure is always recovered from any arbitrary product of boson-mapped spin operators, and that the only deviation from the correct spin algebra arises in the generation of high-order excitations (which are removed upon using the physical basis in the following subsection). This allows one to show that the backward Schwinger transformation of the boson-mapped fermionic operator in Eq.~(\ref{eq:arbitrary-operator-map-schwinger}) is well defined as long as the higher order excitations are removed, i.e., 
\begin{equation} \label{eq:arbitrary-operator-map-schwinger-final-step}
    \hat{O}(\{ \hat{c}_j^{\dagger}, \hat{c}_j \}) \quad \xmapsto[]{JW} \quad \hat{O}(\{ \tilde{\sigma}_j \})  \quad ``\xmapsto[]{Schwinger} " \quad  \hat{O}(\{ \hat{b}_{j\beta}^{\dagger}, \hat{b}_{j\beta}, \hat{b}_{j\alpha}^{\dagger}, \hat{b}_{j\alpha} \}) \quad \xmapsto[]{Schwinger} \quad \hat{O}(\{ \tilde{\sigma}_j \}) + \mathcal{O}(2) + ....
\end{equation}
Equation (\ref{eq:arbitrary-operator-map-schwinger-final-step}) demonstrates that the fermion-to-spin-to-boson transformation presented in this section has a well defined backward transformation, granted that high order excitations are neglected. 
    
A simple example that illustrates this point is the fermionic anticommutation relations after mapping individual creation and annihilation operators to the boson representation using Eqs.~(\ref{eq:JW-transformation-schwinger})-(\ref{eq:definition-nonlocal-F-schwinger}),
\begin{equation}\label{eq:anticommutivity-schwinger-boson}
	\begin{split}
	\{ \hat{c}_j, \hat{c}_k^{\dagger}\} &\mapsto \{ \hat{F}(0, j) \hat{b}_{j\alpha}^{\dagger}\hat{b}_{j\beta}, \hat{F}(0,k) \hat{b}_{k\beta}^{\dagger}\hat{b}_{k\alpha} \} \\
	&= \delta_{j,k} \hat{F}(0, j)^2 \Big[ \hat{b}_{j\alpha}^{\dagger}\hat{b}_{j\beta}\hat{b}_{j\beta}^{\dagger}\hat{b}_{j\alpha} + \hat{b}_{j\beta}^{\dagger}\hat{b}_{j\alpha}\hat{b}_{j\alpha}^{\dagger}\hat{b}_{j\beta} \Big] + (1 - \delta_{j,k}) \hat{F}(0,\min[j,k])^2 \hat{F}(j,k) \\
	&\qquad \times \Bigg[   \hat{b}_{j\alpha}^{\dagger}\hat{b}_{j\beta}\Big(\theta(j-k)(\hat{b}_{k\alpha}^{\dagger}\hat{b}_{k\alpha} -  \hat{b}_{k\beta}^{\dagger}\hat{b}_{k\beta}) + \theta(k-j)(\hat{b}_{j\alpha}^{\dagger}\hat{b}_{j\alpha} -  \hat{b}_{j\beta}^{\dagger}\hat{b}_{j\beta}) \Big)\hat{b}_{k\beta}^{\dagger}\hat{b}_{k\alpha}  \\
	&\qquad \qquad + \hat{b}_{k\beta}^{\dagger}\hat{b}_{k\alpha} \Big(\theta(j-k)(\hat{b}_{k\alpha}^{\dagger}\hat{b}_{k\alpha} -  \hat{b}_{k\beta}^{\dagger}\hat{b}_{k\beta}) + \theta(k-j)(\hat{b}_{j\alpha}^{\dagger}\hat{b}_{j\alpha} -  \hat{b}_{j\beta}^{\dagger}\hat{b}_{j\beta}) \Big) \hat{b}_{j\alpha}^{\dagger}\hat{b}_{j\beta} \Bigg] \\
	&= \delta_{j,k} \hat{F}(0, j)^2 \Big[ \mathbf{1}_j +\mathcal{O}_j(2) \Big] + (1 - \delta_{j,k}) \hat{F}(0,\min[j,k])^2 \hat{F}(j,k) \\
	& \qquad \times \Bigg[ \theta(j-k)\Big[\hat{b}_{j\alpha}^{\dagger}\hat{b}_{j\beta}(-\hat{b}_{k\beta}^{\dagger}\hat{b}_{k\alpha} + \mathcal{O}_k(2)) + (\hat{b}_{k\beta}^{\dagger}\hat{b}_{k\alpha} + \mathcal{O}_k(2))\hat{b}_{j\alpha}^{\dagger}\hat{b}_{j\beta} \Big] \\
	&\qquad \qquad + \theta(k-j)\Big[(-\hat{b}_{j\alpha}^{\dagger}\hat{b}_{j\beta} + \mathcal{O}_j(2))\hat{b}_{k\beta}^{\dagger}\hat{b}_{k\alpha} + (\hat{b}_{j\alpha}^{\dagger}\hat{b}_{j\beta} + \mathcal{O}_j(2))\hat{b}_{k\beta}^{\dagger}\hat{b}_{k\alpha} \Big] \Bigg] \\
	&= \delta_{j,k} \hat{F}(0, j)^2 \Big[ \mathbf{1}_j +\mathcal{O}_j(2) \Big]  + (1 - \delta_{j,k}) \hat{F}(0,\min[j,k])^2 \hat{F}(j,k) \Big[ \theta(j-k)\hat{a}_j^{\dagger}\hat{b}_j\mathcal{O}_k(2)  + \theta(k-j) \hat{b}_k^{\dagger}\hat{a}_k\mathcal{O}_j(2) \Big], \\
	\end{split} 
\end{equation}	
where, choosing $f(\sigma_j^z) = -\sigma_j^z$ for simplicity and using Eq.~(\ref{eq:sigmaz-schwinger-powers}), it is clear that
\begin{subequations}
	\begin{eqnarray}
	\hat{F}(j,k)^{2n} &=& \prod_{l = \min[j,k]}^{\max[j,k]} \Big[\mathbf{1}_l + \mathcal{O}_l(2) + ... + \mathcal{O}_l(2n) \Big], \\
	\hat{F}(j,k)^{2n+1} &=& \prod_{l = \min[j,k]}^{\max[j,k]} \Big[-\sigma^z_l + \mathcal{O}_l(2) + ... + \mathcal{O}_l(2n+1) \Big].
	\end{eqnarray}
\end{subequations}
Importantly, the last line of Eq.~(\ref{eq:anticommutivity-schwinger-boson}) reduces to the fermionic anticommutation relations, $\{ \hat{c}_j, \hat{c}_k^{\dagger}\} = \delta_{j,k}$, in the single-excitation limit, i.e., when $\mathcal{O}(n) \rightarrow 0$ for $n \geq 2$. Following the same approach shown in this example to interrogate the bosonic representation of more complex products of fermionic operators, one recovers the correct spin representation at the single-excitation level with additional high-order excitations. As shown in the next subsection, these unphysical excitations are eliminated when using the physical basis.

\subsection{Removal of unphysical excitations in the fermion-to-boson map via use of the physical basis}
\label{ssec:removal-of-excitations}

Here we show that the spurious high-order excitation operators generated as a consequence of the fermion-to-boson transformation in Sec.~\ref{ssec:unphysical-excitations} can be rigorously removed by using the physical basis in the evaluation of matrix elements. Specifically, it is easy to confirm that, while the fermion-to-boson map in Eq.~(\ref{eq:arbitrary-operator-map-schwinger}) is not exact on the operator level due to its generation of unphysical, high-order excitations, use of the physical basis renders the map an exact isomorphism at the matrix element level, 
\begin{equation} \label{eq:arbitrary-operator-map-schwinger-matrix-elements-si}
    \bra{ \mathbf{\tilde{n}} } \hat{O}(\hat{c}_j, \hat{c}_j^{\dagger})\ket{\mathbf{\tilde{n}}'}  \quad \xmapsto[]{JW} \quad \bra{ \mathbf{\bar{n}} } \hat{O}(\{ \tilde{\sigma}_j \}) \ket{\mathbf{\bar{n}}'}   \quad \xmapsto[]{Schwinger} \quad  \bra{ \mathbf{n} }\hat{O}(\{ \hat{b}_{j\beta}^{\dagger}, \hat{b}_{j\beta}, \hat{b}_{j\alpha}^{\dagger}, \hat{b}_{j\alpha} \}) \ket{\mathbf{n}'},
\end{equation}
where the lack of the quotation marks around the map symbol denotes that the map is exact for the evaluation of matrix elements. Here the physical basis in the boson representation is determined by the fermionic basis. To establish this relationship, one can transform the fermionic many-body basis $\ket{\mathbf{\tilde{n}}}$, to the analogous spin basis, $\ket{\mathbf{\bar{n}}}$, and finally into the appropriate boson basis $\ket{\mathbf{n}}$. Here, $\mathbf{\tilde{n}} \equiv \{ \tilde{n}_1, \tilde{n}_2, ..., \tilde{n}_M \}$ is the \textit{ordered} set of fermion occupation numbers, $\mathbf{\bar{n}} \equiv \{ \bar{n}_1, \bar{n}_2, ..., \bar{n}_M \}$ is the (not necessarily ordered) set of spin ``occupation'' numbers,  and $\mathbf{n} \equiv \{ n_{1\beta}, n_{1\alpha}, n_{2\beta}, n_{2\alpha}, ..., n_{M\beta}, n_{M\alpha} \}$ is the (not necessarily ordered) set of analogous boson occupation numbers.  

Because of the central role that the physical basis plays in the present transformation, its origin merits further consideration. The physical basis originates from the consideration, first, of the Schwinger representation of spin 1/2 operators as coupled bosons. A spin, like the single-particle orbital in a fermionic problem, has a Hilbert space with only two states. For simplicity, and because of the intuitive connection to possible states of a fermionic single-particle orbital, we label these two states as occupied and unoccupied. To identify these two states in the Schwinger boson representation, which fixes the physical basis, it is sufficient to consider the expression for the unit operator in the Schwinger representation, Eq.~(\ref{eq:spin-unit-schwinger-si}). Being the unit operator, when one evaluates its diagonal matrix elements in the physical basis, one must recover the value $1$. Hence,
\begin{equation}
    1 = \bra{\tilde{n}_j}\mathbf{1}_j\ket{\tilde{n}_j}  \quad \xmapsto[]{Schwinger} \quad \bra{n_{j\beta}, n_{j\alpha}}\hat{b}^{\dagger}_{j\beta}\hat{b}_{j\beta}\ket{n_{j\beta}, n_{j\alpha}} + \bra{n_{j\beta}, n_{j\alpha}}\hat{b}^{\dagger}_{j\alpha}\hat{b}_{j\alpha}\ket{n_{j\beta}, n_{j\alpha}} = 1,
\end{equation}
which must hold whether the spin, and therefore fermionic single-particle orbital, is in its occupied or unoccupied state, $\tilde{n}_j = 1,0$. Setting $\tilde{n}_j = 1$ requires one to choose which of the two boson modes, $\alpha$ or $\beta$, should mirror the fermionic occupation number. Here, we have chosen the $\beta$ mode to reflect the fermionic occupation number, which requires that the $\alpha$ mode have the opposite behavior, i.e., that its occupation be anticorrelated to the fermionic occupation of the single-single orbital. This choice dictates that the relationship between the fermionic and bosonic occupation numbers corresponding to a single-particle orbital index, $j$, is given by $n_{j\beta} = \tilde{n}_j$ and $n_{j \alpha} = 1 - \tilde{n}_j$, where $\tilde{n}_j \in \{ 0, 1 \}$. Thus, these considerations can be used to clarify the origins of the physical basis in which one obtains two coupled bosons, $\alpha$ and $\beta$, with anticorrelated excitation numbers constrained to the single-excitation subspace. 

An important consequence of the physical basis arises when considering the transition to Cartesian phase space. Once one has obtained the mapped bosonic Hamiltonian using the mapping procedure outlined Eqs.~(\ref{eq:JW-transformation-schwinger-si})-(\ref{eq:definition-nonlocal-F-schwinger-si}), one can formulate the problem in phase space by replacing the bosonic creation and annihilation operators with their phase space counterparts given in Eq.~(\ref{eq:bosons-as-cartesian}). For manipulations in Cartesian phase space, it is necessary to use the resolution of the identity in continuous space to obtain expressions that can be evaluated using trajectories, such as in path integral, quantum-classical and semiclassical approaches. However, the expression for the resolution of the identity in phase space for bosons in this transformation needs to be modified to reflect that only the ground and first excited states are allowed,
\begin{subequations}
\begin{align}
    \mathbf{1}_{j\gamma} &= \int dq_{j\gamma}\ \mathcal{P}_{j\gamma} \ket{q_{j\gamma}}\bra{q_{j\gamma}} \mathcal{P}_{j\gamma}, \label{eq:resolution-x} \\
    &= \int dp_{j\gamma}\ \mathcal{P}_{j\gamma} \ket{p_{j\gamma}}\bra{p_{j\gamma}} \mathcal{P}_{j\gamma}, \label{eq:resolution-p}
\end{align}
\end{subequations}
where 
\begin{equation}
    \mathcal{P}_{j\gamma} = \sum_{n = 0}^{1}\ket{n_{j\gamma}}\bra{n_{j\gamma}}
\end{equation}
is the projection operator onto the physical subspace, consisting of the ground and first excited harmonic oscillator energy eigenstates, of the $j^{\mathrm{th}}$ $\gamma$-mode boson and $\gamma \in \{ \alpha, \beta \}$. Here it is worth noting the similarity between this Hilbert space restriction and that which must be imposed when working with spin coherent states \cite{Klauder1997-sup} and when using the MMST transformation to map discrete states for the path integral treatment of nonadiabatic problems \cite{Ananth2010-sup, Ananth2013-sup}. This constraint on the Hilbert space ensures that use of the resolution of the identity does not introduce errors associated with excursions into the unphysical space of higher excitations of the mapped bosons.  

Since the matrix elements of high-order excitations, i.e., operators that contain multiple creation or annihilation operators corresponding to the same mode, exactly disappear when using the physical basis defined above, $\bra{ \mathbf{n} }\mathcal{O}(n \geq 2) \ket{\mathbf{n}'} = 0$,
it is clear that Eq.~(\ref{eq:arbitrary-operator-map-schwinger-matrix-elements-si}) must be valid. Thus, \emph{the fermion-to-boson map in Eqs.~(\ref{eq:JW-transformation-schwinger-si}) and (\ref{eq:definition-nonlocal-F-schwinger-si}) is strictly exact when the basis used for the evaluation of matrix elements and traces is restricted to the single-excitation subspace per fermion index $j$}. 

\section{Alternative spin mapping approaches}
\label{sec:alternative-spin-to-boson-maps}

In this section, we analyze the feasibility of using alternative spin-to-boson maps, namely the Holstein-Primakoff \cite{Holstein1940-sup} and Matsubara-Matsuda \cite{Matsubara1956-sup} transformations, for the derivation of fermion-to-Cartesian phase space maps. In particular, we show that, while the Holstein-Primakoff transformation can generally be used in lieu of the Schwinger theory of angular momentum, it introduces undesirable nonlinearities in the form of the square root of the shifted occupation number operator, $(1 - \hat{b}^{\dagger}\hat{b})^{1/2}$. In the case of the Matsubara-Matsuda transformation, we first demonstrate that, when applied to the mapping of discrete states, it yields the MMST transformation. However, we then show that the Matsubara-Matsuda transformation \emph{cannot} be used to obtain an exact means of mapping fermionic creation and annihilation operators to bosonic ones (and, consequently, to phase space variables, using Eq.~(\ref{eq:bosons-as-cartesian})) when combined with the JW transformation. Nevertheless, we demonstrate how it can be used to obtain an \emph{approximate} fermion-to-boson map and analyze how such a map could be used with appropriate constraints to investigate the quantum statics and dynamics of the many-fermion problem.

\subsection{Holstein-Primakoff representation}
\label{ssec:holstein-primakoff}

In the Holstein-Primakoff transformation \cite{Holstein1940-sup}, one uses a single boson for each spin index $j$, by expressing the $\alpha$ boson in the Schwinger mapping in terms of the $\beta$ boson \cite{Mattis1988-sup} by using the completeness relation of the joint Hilbert space, Eq.~(\ref{eq:spin-unit-schwinger-si}),
\begin{subequations}
    \begin{align}
	\sigma_j^+ &\quad ``\xmapsto{HP}" \quad \hat{b}_{j}^{\dagger} (1 - \hat{b}_{j}^{\dagger}\hat{b}_{j})^{1/2}, \\
	\sigma_j^- &\quad ``\xmapsto{HP}" \quad (1 - \hat{b}_{j}^{\dagger}\hat{b}_{j})^{1/2} \ \hat{b}_{j},
	\end{align}
\end{subequations}
which implies that 
\begin{subequations}
	\begin{align}
	\sigma_j^z & \quad``\xmapsto{HP}" \quad 2\hat{b}_j^{\dagger}\hat{b}_j - 1,\\
	\mathbf{1}_j & \quad``\xmapsto{HP}" \quad 1 + \mathcal{O}_j(2).
	\end{align}
\end{subequations}
	
It is straightforward to confirm that the Holstein-Primakoff transformation, like the Schwinger representation, exactly recovers the spin commutation relations, Eq.~(\ref{eq:spin-commutation}) on the operator level, but recovers the spin anticommutation relations, Eq.~(\ref{eq:spin-anticommutation}), accompanied by unphysical excitations. Also similar to the Schwinger representation, the Holstein-Primakoff transformation recovers the correct spin algebra on the matrix element level when the Hilbert space of the mapped bosons is restricted to the subspace consisting of the ground and first excited states of each boson. Hence, following the same argument made in the context of the Schwinger representation in Sec.~\ref{ssec:removal-of-excitations}, the fermion-to-boson (and then to phase space variables) map obtained using the JW transformation followed by the Holstein-Primakoff transformation is exact on the matrix element level when the physical basis is used to evaluate matrix elements and traces. Hence, this allows us to derive a fermion-to-boson and therefore Cartesian phase space operator transformation, which is exact on the matrix element level, that takes the form, 
\begin{subequations} \label{eq:JW-transformation-hp}
    \begin{align}
    	\hat{c}_j  \quad  &``\mapsto " \quad  \hat{F}(0,j) \ (1 - \hat{b}_j^{\dagger}\hat{b}_j)^{1/2}\ \hat{b}_{j} = \frac{1}{2}\hat{F}(0,j)(3 - \hat{q}_j^2 - \hat{p}_j^2)^{1/2}\ (\hat{q}_{j} + i\hat{p}_{j}), \label{eq:JW-transformation-annihilation-hp} \\
    	\hat{c}_j^{\dagger}  \quad &``\mapsto " \quad  \hat{F}(0,j)\  \hat{b}^{\dagger}_{j}\ (1 - \hat{b}_j^{\dagger}\hat{b}_j)^{1/2} = \frac{1}{2}\hat{F}(0,j)(3 - \hat{q}_j^2 - \hat{p}_j^2)^{1/2}\ (\hat{q}_{j} - i\hat{p}_{j}), \label{eq:JW-transformation-creation-hp}
	\end{align}
\end{subequations}
where
\begin{equation}
    \hat{F}(j,k)  \quad ``\mapsto " \quad  \prod_{l=\min[j+1,k+1]}^{\max[j-1,k-1]} f(2\hat{b}^{\dagger}_{j}\hat{b}_{j} - 1) = \prod_{l=\min[j+1,k+1]}^{\max[j-1,k-1]} f(\hat{q}_j^2 + \hat{p}_j^2 - 2). \label{eq:definition-nonlocal-F-hp}
\end{equation}
In this case, the physical basis in the boson representation corresponds to the physical basis of the $\beta$ mode in transformation obtained using the Schwinger representation, i.e., $\tilde{n}_{j} = n_{j}$, where $\tilde{n}_j \in \{0,1\}$ is the fermionic occupation number for the $j^{\mathrm{th}}$ single-particle orbital. For the Cartesian representation of the transformation in Eqs.~(\ref{eq:JW-transformation-hp})-(\ref{eq:definition-nonlocal-F-hp}), the same considerations regarding the restriction on the Hilbert space arising from the projected resolution of the identity in Sec.~\ref{ssec:removal-of-excitations} apply.  

The presence of the square root in the Holstein-Primakoff transformation, however, can lead to complications, especially when combining it with an approximate treatment of the resulting Hamiltonian. Because of this, the square root term is often expanded in terms of the bosonic occupation number operator $\hat{b}_j^{\dagger}\hat{b}_j$, 
\begin{equation}
	(1 - \hat{b}_j^{\dagger}\hat{b}_j)^{1/2} = 1 - \frac{1}{2} \hat{b}_j^{\dagger}\hat{b}_j - \frac{1}{8} [\hat{b}_j^{\dagger}\hat{b}_j]^2 - ...,
\end{equation}	 
and then truncated at some low order, as in the theory of spin waves \cite{Auerbach1998-sup}. However, truncation of this expansion renders the resulting transformation approximate. Furthermore, use of the Holstein-Primakoff representation can lead to difficulties in, for example, semiclassical treatments that require the Wigner transform of the Hamiltonian. Also, as previously noted in the context of semiclassical and quantum-classical treatments of  discrete states mapped using the Holstein-Primakoff transformation \cite{Thoss1999-sup}, it is common for the dynamics (whether in real or imaginary time) to move the system out of the physical subspace, leading to imaginary contributions to the classical limit of the mapped Hamiltonian. However, despite these possible complications, we do not discount cases where the fermion-to-boson (and, consequently, to phase space variables) transformation provided in Eqs.(\ref{eq:JW-transformation-hp})-(\ref{eq:definition-nonlocal-F-hp}) may prove advantageous. 
	
\subsection{Matsuda-Matsubara representation}
\label{ssec:matsubara-matsuda}
		
A different representation of spins, called the Matsubara-Matsuda transformation \cite{Matsubara1956-sup}, uses hard-core bosons, whose Hilbert space is restricted to the zero and one excitation subspace, 
\begin{subequations} \label{eq:sigmas-matsubara-matsuda}
    \begin{align}
    \sigma_j^+ \quad &\xmapsto[]{Mat.-Mat.}  \quad \hat{B}^{\dagger}_{j}, \label{eq:sigma+-matsubara-matsuda} \\
    \sigma_j^- \quad &\xmapsto[]{Mat.-Mat.}  \quad \hat{B}_{j}, \label{eq:sigma--matsubara-matsuda} \\
    \mathbf{1}_j \quad &\xmapsto[]{Mat.-Mat.}  \quad \hat{B}^{\dagger}_{j}\hat{B}_{j} + \hat{B}_{j}\hat{B}_{j}^{\dagger} = 1,\label{eq:spin-unit-matsubara-matsuda}\\
    \sigma_j^{z} \quad &\xmapsto[]{Mat.-Mat.}  \quad \hat{B}^{\dagger}_{j}\hat{B}_{j} - \hat{B}_{j}\hat{B}_{j}^{\dagger} = 2\hat{B}^{\dagger}_{j}\hat{B}_{j} -1, \label{eq:sigmaz-matsubara-matsuda}
    \end{align}
\end{subequations}
where the hard-core constraint limits the Hilbert space of a particular mode to the subspace consisting of zero and one excitations. This restriction of the Hilbert space can be captured via the commutation relation 
\begin{equation}\label{eq:hc-boson-commutation-relations}
    [\hat{B}_j, \hat{B}_k^{\dagger}] = \delta_{j,k}(1 - 2 \hat{B}_j^{\dagger}\hat{B}_j).
\end{equation}
In other words, in contrast to the Schwinger and Holstein-Primakoff transformations, this transformation is exact at the operator level, since different bosonic modes commute, while the creation and annhilation operators of a particular hard-core bosonic mode anticommute $\{ \hat{B}_j, \hat{B}_j^{\dagger} \} = 1$. 
    
\subsubsection{Application to $N$-level systems: Relation to the MMST transformation}
\label{sssec:matsubara-matsuda-regular-boson-mmst}

Before we consider its use in the development of a fermionic map, it is informative to consider how the Matsubara-Matsuda transformation can be understood as the generator of the MMST transformation. To appreciate this, we note that an arbitrary $N$-level system outer product, $\ket{j}\bra{k}$, can be replaced by a two-body product of spin operators that can then be mapped to hard-core bosons using the Matsubara-Matsuda transformation,
\begin{subequations}\label{eq:matsubara-matsuda-for-mmst-hc}
\begin{align}
    \ket{j}\bra{k} &\mapsto \sigma_j^{+} \sigma_k^{-} = \ket{1_j, 0_k}\bra{0_j, 1_k}, \\
    &\mapsto \hat{B}_j^{\dagger}\hat{B}_k.
\end{align}
\end{subequations}
Thus, the the physical basis for the $N$-level system can be expressed as a many-body boson basis subject to two restrictions. The first is that of hard-core bosons themselves, which requires that for each \emph{individual} hard-core boson, the physical Hilbert space spans only the zero and one excitation subspace. The second constraint corresponds to the fact that the physical basis for the $N$-level system translates to the \emph{collective} one-excitation manifold of all possible many-body hard-core boson states, i.e., the set of many-body states where one hard-core boson is in the first excited state while the rest remain in their ground states. 

To obtain a phase space description of $N$-level systems, one would ideally want to use the relationship between regular bosonic creation and annihilation operators and continuous position and momentum operators, Eq.~(\ref{eq:bosons-as-cartesian}), for hard-core boson operators. Such a replacement, however, would yield operators that are unable to recover the hard-core boson commutation relations, Eq.~(\ref{eq:hc-boson-commutation-relations}). While such a replacement is not exact on the operator level, use of the physical basis leads to an exact isomorphism at the matrix-element level. Thus, one can replace the hard-core bosons $\{\hat{B}_j^{\dagger}, \hat{B}_j \}$ in Eq.~(\ref{eq:matsubara-matsuda-for-mmst}) with regular bosons, $\hat{b}_j^{\dagger}, \hat{b}_j$,  yielding the MMST transformation, 
\begin{equation} \label{eq:matsubara-matsuda-for-mmst}
	\ket{j}\bra{k} \quad ``\mapsto" \quad \hat{b}_j^{\dagger}\hat{b}_k.
\end{equation}
Here, like in the Schwinger boson representation of spin operators, we emphasize that the map is \emph{not} exact at the operator level, as can be confirmed via the anticommutation relation of $N$-level system outer products, 
\begin{equation}
    \begin{split}
    \{ \ket{j}\bra{k}, \ket{m}\bra{n} \} \quad ``&\mapsto" \quad \{ \hat{b}_j^{\dagger}\hat{b}_k, \hat{b}^{\dagger}_{m}\hat{b}_n \}  =  \delta_{km} \hat{b}_j^{\dagger}\hat{b}_n + \delta_{nj} \hat{b}_m^{\dagger}\hat{b}_k + 2 \hat{b}^{\dagger}_{j}\hat{b}^{\dagger}_{m}\hat{b}_n\hat{b}_k \\
    &\mapsto \quad \ \ \delta_{km} \ket{j}\bra{n} + \delta_{nj} \ket{m}\bra{k} + \mathcal{O}(2).
    \end{split}
\end{equation}
However, the map becomes exact at the matrix element level when it is used in conjunction with the physical basis to obtain the matrix elements of, for example, the propagator and, more broadly, of an arbitrary operator \cite{Stock1997-sup}, 
\begin{equation} \label{eq:mmst-matrix-level}
    \braket{n| \hat{O}(\{ \ket{j}\bra{k} \})|m} \mapsto \braket{0_1, ..., 1_n, ..., 0_m, ..., 0_N | \hat{O}(\{ \hat{b}_{j}^{\dagger}\hat{b}_{k} \}) | 0_1, ..., 0_n, ..., 1_m, ..., 0_N }.
\end{equation}
Hence, \emph{while the MMST map is not exact on an operator level, it exactly reproduces the matrix structure when used in conjunction with the physical basis.} 

\subsubsection{Application to many-fermion problems}
\label{sssec:matsubara-matsuda-regular-boson-application-to-many-fermion-problems}
    
Considering the success of the MMST transformation in the treatment of nonadibatic problems, it is desirable to consider whether the Matsubara-Matsuda transformation \cite{Matsubara1956-sup} followed by the relaxation of the hard-core constraint can also be used to obtain an exact fermion-to-boson map. 

We begin by substituting the Matsubara-Matsuda expression of spin operators, Eq.~(\ref{eq:sigmas-matsubara-matsuda}) into the JW transformed expression for fermionic operators, Eqs.~(\ref{eq:JW-transformation})-(\ref{eq:definition-nonlocal-F}), yielding, 
\begin{subequations} \label{eq:JW-transformation-matsubara-matsuda}
    \begin{align}
    	\hat{c}_j \quad &\mapsto \quad  \hat{F}(0,j) \hat{B}_{j}, \label{eq:JW-transformation-annihilation-matsubara-matsuda} \\
    	\hat{c}_j^{\dagger} \quad &\mapsto \quad  \hat{F}(0,j) \hat{B}^{\dagger}_{j}, \label{eq:JW-transformation-creation-matsubara-matsuda}
    \end{align}
\end{subequations}
where
\begin{equation}
    \hat{F}(j,k) \quad \mapsto \quad  \prod_{l=\min[j+1,k+1]}^{\max[j-1,k-1]} f(2 \hat{B}^{\dagger}_{j}\hat{B}_{j}-1). \label{eq:definition-nonlocal-F-matsubara-matsuda}
\end{equation}
The isomorphism between hard-core bosons and spins establishes the exact nature of the above map. However, our goal is an expression for fermionic operators in terms of continuous positions and momenta. To that end, it is desirable to replace the hard-core bosons with regular bosons, whose algebra can be easily connected with that of phase space variables, and assess the validity of the resulting transformation. Unfortunately, in contrast to the MMST transformation, we now show that the resulting map is \emph{not} valid at the operator or matrix element level. In other words,
\begin{subequations} 
    \begin{align} 
    &\quad\ \ \hat{O}(\hat{c}_j, \hat{c}_j^{\dagger})  &&\xmapsto[]{JW} &&\quad\ \  \hat{O}(\{ \tilde{\sigma}_j \})  && \xmapsto[]{Mat.-Mat.}  &&\quad\ \  \hat{O}(\{ \hat{B}_{j}^{\dagger}, \hat{B}_{j}\})   &&\centernot \mapsto &&\quad\ \   \hat{O}(\{ \hat{b}_{j}^{\dagger}, \hat{b}_{j}\}) ,  \label{eq:arbitrary-operator-map-matsubara-matsuda-operator-level} \\
    & \bra{ \mathbf{\tilde{n}} } \hat{O}(\hat{c}_j, \hat{c}_j^{\dagger}) \ket{\mathbf{\tilde{n}}'} &&\xmapsto[]{JW}  && \bra{ \mathbf{\bar{n}} } \hat{O}(\{ \tilde{\sigma}_j \}) \ket{\mathbf{\bar{n}}'} && \xmapsto[]{Mat.-Mat.}  && \bra{ \mathbf{n} } \hat{O}(\{ \hat{B}_{j}^{\dagger}, \hat{B}_{j}\}) \ket{\mathbf{n}'} && \centernot \mapsto  && \bra{ \mathbf{n} }\hat{O}(\{ \hat{b}_{j}^{\dagger}, \hat{b}_{j}\}) \ket{\mathbf{n}'}.  \label{eq:arbitrary-operator-map-matsubara-matsuda-matrix-elements}
    \end{align}
\end{subequations}
Here, the physical basis of the mapped bosons, $\{ \hat{b}_j^{\dagger}, \hat{b}_j\}$, is equivalent to that of the $\beta$ boson in the fermion-to-boson transformation presented in Sec.~(\ref{ssec:removal-of-excitations}), where  $\tilde{n}_{j} = n_{j}$, where $\tilde{n}_j \in \{0,1\}$ is the fermionic occupation number for the $j^{\mathrm{th}}$ single-particle orbital. We illustrate the inability of the resulting transformation to treat the many-fermion case by considering one of simplest products of mapped fermionic operators. First, we consider the expected fermionic result,
\begin{equation}\label{eq:matusbara-matsuda-failure-fermion-matrix-element}
    \begin{split}
    \bra{ \mathbf{\tilde{n}} } \hat{c}_j\hat{c}_k^{\dagger}  \ket{\mathbf{\tilde{n}}'} &=\bra{ \mathbf{\tilde{n}} } (\delta_{j,k} - \hat{c}_k^{\dagger}\hat{c}_j)  \ket{\mathbf{\tilde{n}}'}\\
    &= \delta_{j,k} g(\tilde{\mathbf{n}}, \tilde{\mathbf{n}}') (1 - \delta_{\tilde{n}'_j, 1}) - (1 - \delta_{j,k}) g(\tilde{\mathbf{n}}, \tilde{\mathbf{n}}' | j,k) h(\tilde{\mathbf{n}}'; j,k),
    \end{split}
\end{equation}
where
\begin{align}
	g(\tilde{\mathbf{n}},\tilde{\mathbf{n}}') &= \prod_{r = 1}^{M} \delta_{\tilde{n}_r,\tilde{n}_r'}, \\
	g(\tilde{\mathbf{n}},\tilde{\mathbf{n}}'|\mathbf{m}) &= \prod_{\substack{r = 1\\ r \neq m_1, m_2, ...}}^{M} \delta_{\tilde{n}_r,\tilde{n}_r'},\\
	h(\tilde{\mathbf{n}};\mathbf{j}) &= \Bigg[\prod_{r = s_1+1}^{s_2-1} (-1)^{\tilde{n}_r} \Bigg] \Bigg[\prod_{r = s_3+1}^{s_4-1} (-1)^{\tilde{n}_r}\Bigg] \times ... ,\\
\end{align}
and $s_1,s_2,...$ correspond to $j_1, j_2,...$ rearranged in order of increasing magnitude.
	
Now we consider the matrix elements of the mapped version,
\begin{equation}\label{eq:matusbara-matsuda-failure-mapped-fermion-matrix-element}
	\begin{split}
    \bra{ \mathbf{n} } \hat{F}(0,j)\hat{b}_j\hat{F}(0,k)\hat{b}_k^{\dagger} \ket{\mathbf{n}'} &= \delta_{j,k}  \bra{ \mathbf{n} } \hat{F}(0,j)^2 (1 + \hat{b}_j^{\dagger}\hat{b}_j)\ket{\mathbf{n}'} + (1- \delta_{j,k})  \bra{ \mathbf{n} } \hat{F}(0,\min[j,k])^2 \hat{F}(j,k) \\
     &\qquad \qquad \qquad \qquad \qquad \times \Big[ \theta(j - k) \hat{b}_j(1 - 2\hat{b}_{k}^{\dagger}\hat{b}_{k})\hat{b}_k^{\dagger} + \theta(k-j) \hat{b}_j(1 - 2\hat{b}_{j}^{\dagger}\hat{b}_{j})\hat{b}_k^{\dagger} \Big]\ket{\mathbf{n}'}  \\  
    &= \delta_{j,k}  g(\mathbf{n}, \mathbf{n}') (1 + \delta_{n'_j, 1}) + (1- \delta_{j,k})  \bra{ \mathbf{n} } \hat{F}(0,\min[j,k])^2 \hat{F}(j,k) \\
    &\qquad \qquad \qquad \qquad \qquad \times \Big[\hat{b}_j\hat{b}_k^{\dagger}  -2\theta(j - k) \hat{b}_j\hat{b}_{k}^{\dagger}\hat{b}_{k}\hat{b}_k^{\dagger} -2\theta(k-j) \hat{b}_j\hat{b}_{j}^{\dagger}\hat{b}_{j}\hat{b}_k^{\dagger} \Big]\ket{\mathbf{n}'}  \\
    &= \delta_{j,k}  g(\mathbf{n}, \mathbf{n}') (1 + \delta_{n'_j, 1}) \\
    &\qquad \qquad + (1- \delta_{j,k})  \bra{ \mathbf{n} } \hat{F}(0,\min[j,k])^2 \hat{F}(j,k)\Big[-\hat{b}_k^{\dagger}\hat{b}_j  + \mathcal{O}_k(2) + \mathcal{O}_j(2) \Big]\ket{\mathbf{n}'}  \\
    &= \delta_{j,k}  g(\mathbf{n}, \mathbf{n}') (1 + \delta_{n'_j, 1})  - (1- \delta_{j,k})  g(\mathbf{n}, \mathbf{n}' | j,k) h(\mathbf{n}'; j,k),
    \end{split}
\end{equation}
where we have used the fact that
\begin{equation}
	\begin{split}
	\hat{F}(0, n)^2 &= \prod_{l = 1}^{l = n-1} (1 - 2 \hat{b}_l^{\dagger}\hat{b}_l)^2\\
	&= \prod_{l = 1}^{l = n-1} (1 + 4 \hat{b}_l^{\dagger}\hat{b}_l^{\dagger}\hat{b}_l\hat{b}_l)\\
	&= 1 + \mathcal{O}(2) + ... + \mathcal{O}(2n - 2),
	\end{split}
\end{equation}
and where the physical basis assigned to the bosons $\mathbf{n}$ is equivalent to the fermion basis $\tilde{\mathbf{n}}$, i.e. $\{ \tilde{n}_1 = n_1, ..., \tilde{n}_M = n_M \}$. Hence, Eqs.~(\ref{eq:matusbara-matsuda-failure-fermion-matrix-element}) and (\ref{eq:matusbara-matsuda-failure-mapped-fermion-matrix-element}) demonstrate that in general 
\begin{equation}\label{eq:matusbara-matsuda-failure-fermion-mapping}
    \bra{ \mathbf{\tilde{n}} } \hat{O}(\{\hat{c}_j^{\dagger}, \hat{c}_j\})  \ket{\mathbf{\tilde{n}}'} \quad\centernot \mapsto \quad \bra{ \mathbf{n} } \hat{O}(\{\hat{b}_j^{\dagger}, \hat{b}_j\}) \ket{\mathbf{n}'}.
\end{equation}
Hence, \emph{the Matsubara-Matsuda mapping, which lies at the heart of the derivation of the MMST transformation for systems with discrete states, cannot be used to obtain an analogous exact map for fermionic creation and annihilation operators.} 
	
One may understand the failure of the Matubara-Matsuda map combined with the removal of the hard-core constraint to properly capture properties of fermion operators by considering its more fundamental failure to capture the algebra of spins. Effectively, it is impossible to construct the correct expressions for the unit and spin polarization operators, $\mathbf{1}_j$ and $\sigma_j^z$, from the ladder operators, $\sigma_j^+$ and $\sigma_j^-$, in the version of the Matusbara-Matsuda transformation where the hard-core constraint has been removed,  
\begin{subequations} \label{eq:sigmas-matsubara-matsuda-regular-boson}
    \begin{align}
    \sigma_j^+ \quad &\xmapsto[]{}  \quad \hat{b}^{\dagger}_{j}, \label{eq:sigma+-matsubara-matsuda-regular-boson} \\
    \sigma_j^- \quad &\xmapsto[]{}  \quad \hat{b}_{j}, \label{eq:sigma--matsubara-matsuda-regular-boson} \\
    \mathbf{1}_j  \quad &\xmapsto[]{}  \quad \hat{b}^{\dagger}_{j}\hat{b}_{j} + \hat{b}_{j}\hat{b}_{j}^{\dagger} = 2\hat{b}^{\dagger}_{j}\hat{b}_{j} + 1 \neq 1,\label{eq:spin-unit-matsubara-matsuda-regular-boson}\\
    \sigma_j^{z} \quad &\xmapsto[]{}  \quad \hat{b}^{\dagger}_{j}\hat{b}_{j} - \hat{b}_{j}\hat{b}_{j}^{\dagger} \neq 2\hat{b}^{\dagger}_{j}\hat{b}_{j} -1. \label{eq:sigmaz-matsubara-matsuda-regular-boson}
    \end{align}
\end{subequations} 
In contrast, the Schwinger theory of angular momentum allows both the unit and spin polarization operators, $\mathbf{1}_j$ and $\sigma_j^z$, to be constructed from the mapped ladder operators, $\sigma_j^+$ and $\sigma_j^-$, leading to expressions that recover the spin algebra on the (joint) single-excitation manifold of the $\alpha$ and $\beta$ modes (see Eqs.~(\ref{eq:sigmas-schwinger-si})). 

\subsubsection{Controlled use of the fermion-to-boson map in Matsubara-Matsuda representation}
\label{sssec:controlled-use-matsubara-matsuda}

While the Matsubara-Matsuda transformation cannot be used to obtain a rigorous map connecting fermionic and bosonic operators, the resulting transformation can be used safely, granted that certain restrictions are imposed. For example, we note that for operators that contain a ``small-scale" product of fermionic operators, $\hat{O}(\{\hat{c}_j^{\dagger}, \hat{c}_j \})$, one can use the map in Eqs.~(\ref{eq:JW-transformation-matsubara-matsuda})-(\ref{eq:definition-nonlocal-F-matsubara-matsuda}), rearrange the resulting operator product in its normal-ordered form while still in the hard-core boson representation, and then relax the hard-core boson constraint. The resulting fermionic operator, expressed entirely in terms of regular bosonic operators, will recover the exact matrix elements as the original fermionic operator, 
\begin{equation} 
    \bra{ \mathbf{\tilde{n}} } \hat{O}(\hat{c}_j, \hat{c}_j^{\dagger}) \ket{\mathbf{\tilde{n}}'} \ \xmapsto[]{JW} \ \bra{ \mathbf{\bar{n}} } \hat{O}(\{ \tilde{\sigma}_j \}) \ket{\mathbf{\bar{n}}'} \ \xmapsto[]{Mat.-Mat.} \ \bra{ \mathbf{n} } \hat{O}(\{ \hat{B}_{j}^{\dagger}, \hat{B}_{j}\})_{\mathrm{ord}} \ket{\mathbf{n}'} \ \mapsto \ \bra{ \mathbf{n} }\hat{O}(\{ \hat{b}_{j}^{\dagger}, \hat{b}_{j}\})_{\mathrm{ord}} \ket{\mathbf{n}'}. \label{eq:arbitrary-operator-map-matsubara-matsuda-matrix-elements-ordered-form}
\end{equation}
In addition, because it can be expressed exclusively in terms of bosonic operators, it is straightforward to obtain a phase space formulation by rewriting the operators in terms of continuous positions and momenta using Eq.~(\ref{eq:bosons-as-cartesian}). Despite the apparent similarity between Eqs.~(\ref{eq:arbitrary-operator-map-matsubara-matsuda-matrix-elements}) and (\ref{eq:arbitrary-operator-map-matsubara-matsuda-matrix-elements-ordered-form}), there is one critical difference: the mapped operator $\hat{O}(\{ \hat{B}_{j}^{\dagger}, \hat{B}_{j}\})_{\mathrm{ord}}$ is placed in its normal-ordered form (thus the ``ord'' subscript) before the hard-core boson constraint is removed. For operators where the functional form requires an infinite expansion of products of fermionic creation and annihilation operators, such as the real and imaginary time propagators, performing these manipulations term by term in the expansion is impractical and ultimately amounts to solving the fermionic problem in its original complexity, which would eliminate the advantages of having a transformation to continuous phase space variables that could be directly used in path integral, quantum-classical, and semiclassical schemes.

Nevertheless, here we examine the utility of Eq.~(\ref{eq:arbitrary-operator-map-matsubara-matsuda-matrix-elements-ordered-form}) for such operators. Because of the central role of the (imaginary or real time) propagator to most problems of interest, and because its treatment illustrates the concerns that arise when using the regular bosonic form of the transformation Eqs.~(\ref{eq:JW-transformation-matsubara-matsuda})-(\ref{eq:definition-nonlocal-F-matsubara-matsuda}), we specialize our discussion to these operators.

We begin by considering the argument of the propagator, the many-fermion Hamiltonian. First, we arrange the mapped hard-core bosonic Hamiltonian in its normal-ordered form and then relax the hard-core boson constraint, 
\begin{equation}\label{eq:hamiltonian-matsubara-matsuda-cartesian-approximation}
    \hat{H}_{\mathrm{ord}}(\{\hat{B}_{j}^{\dagger}, \hat{B}_{j} \}) \quad ``\mapsto" \quad \hat{H}_{\mathrm{ord}}(\{\hat{b}_{j}^{\dagger}, \hat{b}_{j} \}).
\end{equation}
To illustrate this procedure, consider the general many-fermion Hamiltonian in Eq.~(\ref{eq:general-fermion-hamiltonian}), which we map using to its hard-core boson representation using Eqs.~(\ref{eq:JW-transformation-matsubara-matsuda})-(\ref{eq:definition-nonlocal-F-matsubara-matsuda}) and arrange in its normal-ordered form,
\begin{equation}\label{label:general-fermion-hamiltonian-matsubara-matsuda}
    \hat{H}_{\mathrm{ord}} \mapsto \sum_{j,k} h_{j,k}(\boldsymbol{\Gamma}) \hat{F}(j,k)\hat{B}_{j}^{\dagger}\hat{B}_{k} + \frac{1}{2} \sum_{j,k,l,m} U_{jk,lm}(\boldsymbol{\Gamma})\mathrm{sgn}(j-k) \mathrm{sgn}(l-m) \hat{F}(j,k,l,m) \hat{B}_j^{\dagger}\hat{B}_k^{\dagger}\hat{B}_m \hat{B}_l.
\end{equation}
Once in this form, it is possible to replace the hard-core boson operators $\{\hat{B}_j, \hat{B}_j^{\dagger}\}$ with their regular boson counterparts $\{\hat{b}_j, \hat{b}_j^{\dagger}\}$ and then with their phase space expressions $\{\hat{q}_j, \hat{p}_j^{\dagger}\}$, 
\begin{equation}\label{label:general-fermion-hamiltonian-matsubara-matsuda-cartesian}
    \begin{split}
    \hat{H}_{\mathrm{ord}} \quad &``\mapsto" \quad  \sum_{j,k} h_{j,k}(\boldsymbol{\Gamma}) \hat{F}(j,k)\hat{b}_{j}^{\dagger}\hat{b}_{k} + \frac{1}{2} \sum_{j,k,l,m} U_{jk,lm}(\boldsymbol{\Gamma})\mathrm{sgn}(j-k) \mathrm{sgn}(l-m) \hat{F}(j,k,l,m) \hat{b}_j^{\dagger}\hat{b}_k^{\dagger}\hat{b}_m \hat{b}_l\\
    &= \sum_{j,k} h_{j,k}(\boldsymbol{\Gamma}) \frac{1}{2}\hat{F}(j,k)(\hat{q}_{j} - i\hat{p}_{j})(\hat{q}_{k} + i\hat{p}_{k}) \\
    &\qquad+ \frac{1}{2^3} \sum_{j,k,l,m} U_{jk,lm}(\boldsymbol{\Gamma})\mathrm{sgn}(j-k) \mathrm{sgn}(l-m) \hat{F}(j,k,l,m) (\hat{q}_{j} - i\hat{p}_{j})(\hat{q}_{k} - i\hat{p}_{k})(\hat{q}_{m} + i\hat{p}_{m})(\hat{q}_{l} + i\hat{p}_{l}).
    \end{split}
\end{equation}
where 
\begin{equation} \label{eq:definition-nonlocal-F-matsubara-matsuda-cartesian}
    \hat{F}(j,k) \quad ``\mapsto" \quad \prod_{l=\min[j+1,k+1]}^{\max[j-1,k-1]} f(2\hat{b}_l^{\dagger}\hat{b}_l -1) = \prod_{l=\min[j+1,k+1]}^{\max[j-1,k-1]} f(\hat{q}_{l}^2 + \hat{p}_{l}^2 -2),
\end{equation}
and its multi-index generalization, $\hat{F}(\mathbf{s})$, is given by Eq.~(\ref{eq:multi-index-nonlocal-F}). We again place quotation marks around the map sign in Eqs.~(\ref{eq:hamiltonian-matsubara-matsuda-cartesian-approximation})-(\ref{eq:definition-nonlocal-F-matsubara-matsuda-cartesian}) to emphasize that this action is only valid to obtain the matrix elements of the mapped Hamiltonian, but not for use in additional manipulations at the operator level. We now consider the propagator and note that while the normal-ordered hard-core bosonic form of the propagator can be replaced with its regular bosonic counterpart, the same cannot be said when only the Hamiltonian is in its normal-ordered form
\begin{subequations}
\begin{align}
    \bra{ \mathbf{n} }\Big(\exp\Big[-a\hat{H}(\{\hat{B}_{j}^{\dagger}, \hat{B}_{j} \})\Big]\Big)_{\mathrm{ord}}\ket{\mathbf{n}'} & \quad \mapsto \quad \bra{ \mathbf{n} }\Big(\exp\Big[-a\hat{H}(\{\hat{b}_{j}^{\dagger}, \hat{b}_{j} \})\Big]\Big)_{\mathrm{ord}}\ket{\mathbf{n}'},\\ 
    \bra{ \mathbf{n} }\exp\Big[-a\hat{H}_{\mathrm{ord}}(\{\hat{B}_{j}^{\dagger}, \hat{B}_{j} \})\Big]\ket{\mathbf{n}'} & \quad \centernot \mapsto \quad \bra{ \mathbf{n} }\exp\Big[-a\hat{H}_{\mathrm{ord}}(\{\hat{b}_{j}^{\dagger}, \hat{b}_{j} \})\Big]\ket{\mathbf{n}'}, \label{eq:matsubara-matsuda-propagator-failure}
\end{align}
\end{subequations}
where $a \in \{\beta, \pm it \}$, corresponding to the imaginary and real time propagators, respectively. One can prove that the mapping in Eq.~(\ref{eq:matsubara-matsuda-propagator-failure}) is not exact by noting that the matrix elements of the hard-core and regular bosonic Hamiltonians raised to a power $n \geq 2$ yield different results, 
\begin{equation}\label{eq:hamiltonian-powers-failure}
    \bra{ \mathbf{n} }\Big[\hat{H}_{\mathrm{ord}}(\{\hat{B}_{j}^{\dagger}, \hat{B}_{j} \})\Big]^n\ket{\mathbf{n}'} \neq \bra{ \mathbf{n} }\Big[\hat{H}_{\mathrm{ord}}(\{\hat{b}_{j}^{\dagger}, \hat{b}_{j} \})\Big]^n\ket{\mathbf{n}'}.
\end{equation}
Therefore, the matrix elements of the expansion of the propagator are only exact up to first order in the Hamiltonian. However, since this approach has accuracy to first order, when the (real or imaginary) time, $a$, in the propagator is sufficiently small,
\begin{equation}
    \bra{ \mathbf{n} } \exp\Big[a\hat{H}(\{\hat{B}_{j}^{\dagger}, \hat{B}_{j} \})\Big]\Big)_{\mathrm{ord}}\ket{\mathbf{n}'} \approx \bra{ \mathbf{n} } \Big[ 1 - a\hat{H}_{\mathrm{ord}}(\{\hat{b}_{j}^{\dagger}, \hat{b}_{j} \})\Big]\ket{\mathbf{n}'}
\end{equation}
Given this, it may be possible to use this map by decomposing the propagator in a path integral scheme where one splits the propagator into $P$ time slices and then inserts the projector corresponding to the mapped physical subspace,
\begin{equation}
    \bra{ \mathbf{n} }\Big(\exp\Big[-a\hat{H}(\{\hat{B}_{j}^{\dagger}, \hat{B}_{j} \})\Big]\Big)_{\mathrm{ord}}\ket{\mathbf{n}'} \approx \lim_{P \rightarrow \infty} \prod_{k = 1}^{P} \bra{ \mathbf{n} } \Big(\exp\Big[-(a/P)\hat{H}_{\mathrm{ord}}(\{\hat{b}_{j}^{\dagger}, \hat{b}_{j} \})\Big]\mathcal{P} \Big)^P  \ket{\mathbf{n}'}
\end{equation}
where 
\begin{equation}
    \mathcal{P} = \prod_{j = 1}^{M} \mathcal{P}_j = \prod_{j = 1}^{M} \sum_{n_j = 0}^{1} \ket{n_j}\bra{n_j}.
\end{equation}
This way of restricting the physical subspace in a path integral scheme by imposing a constraint on the resolution of the identity in phase space has been used in the similar problem of deriving path integral expressions for systems with discrete states mapped using the MMST transformation \cite{Ananth2010-sup, Ananth2013-sup}. Moreover, we emphasize that once the many-fermion Hamiltonian is written in terms of bosonic creation and annihilation operators, the transformation to Cartesian positions and momenta only requires substituting these operators with their phase space expressions given in Eq.~(\ref{eq:bosons-as-cartesian}).

Hence, as we have shown above, \emph{although the fermion-to-boson transformation that results from using the JW and regular boson form of the Matusubara-Matsuda transformations does not constitute an isomorphism at the operator or matrix element levels, could potentially be used in a controlled manner within the path integral framework to derive expressions in the short timestep limit that can be used to study the statics and dynamics of many-fermion systems.}

\end{document}